\documentclass[
reprint,
superscriptaddress,
 amsmath,amssymb,
 aps,
prfluids, onecolumn,
nofootinbib]{revtex4-2}

\usepackage[utf8]{inputenc}
\usepackage{amsmath,amssymb,bm}
\usepackage{graphicx,color}
\usepackage{lineno}
\usepackage{soul}
\usepackage{hyperref} 
\hypersetup{
    colorlinks=true,
    linkcolor=blue,
    citecolor=blue,      
    urlcolor=blue
    }

\newcommand{\e}{\exp\left(-\left(n+\frac{1}{2}\right)\eta_{0}\right)}

\newcommand{\grad}{\boldsymbol{\nabla}}

\newcommand\diff{\mathrm{d}}
\newcommand{\mat}[1]{\vec{\hat{\mathcal{#1}}}}
\newcommand{\unitvec}[1]{\boldsymbol{{\hat{#1}}}}

\newcommand{\norm}[1]{\left\lVert #1 \right\rVert}
\renewcommand{\vec}[1]{\boldsymbol{#1}}

\begin{document}


\title{Squirming motion near corrugated surfaces}

\author{Sagnik Garai}
\affiliation{Max Planck Institute for the Physics of Complex Systems, N\"othnitzer Stra{\ss}e 38,
01187 Dresden, Germany}
\author{Thomas G. Fai}
\affiliation{Department of Mathematics, Brandeis University, Waltham MA}
\author{Christina Kurzthaler}
\email{ckurzthaler@pks.mpg.de}
\affiliation{Max Planck Institute for the Physics of Complex Systems, N\"othnitzer Stra{\ss}e 38,
01187 Dresden, Germany}
\affiliation{Center for Systems Biology Dresden, Pfotenhauerstraße 108, 01307 Dresden, Germany}
\affiliation{Cluster of Excellence, Physics of Life, TU Dresden, Arnoldstraße 18, 01062 Dresden, Germany}


\begin{abstract}
 Swimming microorganisms often operate in complex confinement, where an interplay of long-ranged hydrodynamic interactions and a short-ranged repulsive interaction can give rise to interesting dynamical behaviors. Here, we theoretically investigate the trajectories of microswimmers -- modeled as squirmers -- in the presence of periodic boundaries.  The latter modify their swimming velocity, leading to behaviors that differ qualitatively from swimming near planar walls. Using a perturbative approach based on bispherical coordinates and the Lorentz reciprocal theorem, we characterize the interaction between a squirmer and a periodic surface in the limit of small surface amplitude and systematically explore its dependence on the boundary corrugation wavelength, squirmer type, orientation, and swimmer–surface distance. Most importantly, our results reveal that pullers become trapped in the valley of the surface corrugations, in contrast to their sliding motion near planar walls. Furthermore, the near-surface dynamics of pushers display oscillations, reflecting the periodicity of the surface structure. A tilt of the swimmer orientation with respect to the surface corrugations results in a wave-length dependent drift that sorts pushers from pullers. These findings highlight the impact of hydrodynamic interactions in shaping microswimmer transport near structured boundaries with potential implications for microbiological phenomena, such as biofilm formation, and technological applications.
\end{abstract}


\maketitle
\section{Introduction}
Self-propelling agents -- ranging from biological microswimmers, such as bacteria, to synthetic colloids -- exhibit interesting dynamical features in confinement~\citep{Xiao:2019,Granek:2024}, where different physical forces underlie surface interactions. Exemplars include surface accumulation~\citep{Berke:2008, Rothschild:1963, Denissenko:2012} and circular dynamics near planar boundaries\citep{Lauga:2006,Berke:2008,Li:2009,Drescher:2011,Bianchi:2017, Di_Leonardo:2011}, which have been attributed to an interplay of long-ranged hydrodynamic interactions~\citep{Berke:2008} and steric repulsion~\citep{Elgeti:2013, Baouche:2026}. 
Despite the abundance of complex boundaries in biological and microfluidic settings, our understanding of confined active systems is mostly restricted to planar boundaries~\citep{Bechinger:2016, Elgeti:2013} while the impact of surface structure remains largely unexplored. Disentangling the interplay of near-field hydrodynamic interactions, steric repulsion, electrostatic or chemical effects near perturbed boundaries at the single-particle level is expected to shed light into microswimmer transport in real environments and collective phenomena, such as the formation of microbial communities~\citep{Secchi:2020, Pellegrino:2026, Yeo:2026}.

Experimental studies on the interactions of microswimmers with complex surface structures are in their infancy. First experimental work has demonstrated that surface topographies can guide the motion of catalytically-driven Janus colloids~\citep{Simmchen:2016}. Recently, {\it Chlamydomonas reinhardtii} have been shown to spend more time in surface valleys~\citep{Li:2026}, while {\it Escherichi coli} tend to accumulate behind surface hills, in the presence of flows~\citep{Secchi:2020}. Alternatively, insights into fluid-structure interactions near complex boundaries have been obtained by studying passive particles sedimenting near corrugated surfaces~\citep{Chase:2022}. Specifically, passive spheres drift along surface corrugations tilted relative to gravity. Direct comparison of experiments and theory demonstrate that the drift results from the hydrodynamic interaction of the particle with the boundary. This naturally urges the question about the role of hydrodynamics in the dynamics of self-propelling particles near corrugated surfaces and whether the latter can be designed to sort them with respect to their flow fields. 

From a theoretical perspective fluid-structure interactions between microswimmers and nearby boundaries are governed by the Stokes equations, as viscous forces typically dominate inertia, and have been mostly studied for two models: The first describes the flow-fields produced by the microswimmer within a far-field approach, which treats the microswimmer as a point particle and consists of the leading order terms in the multipole expansion of the Stokes solution. In this regime, it was found that the corrugation alters the velocity of the microswimmer~\citep{Kurzthaler:2021}, yet the accuracy of such a model starts breaking down as the swimmer is close to a boundary,  i.e. when its distance from the boundary is smaller than its characteristic size. In this regime, the near-field hydrodynamic effects start dominating, as has been studied well for a microswimmer near a rigid wall using numerical and semi-analytical tools \citep{Schaar:2015,Lintuvuori:2016}. 

The second canonical microswimmer model is the spherical squirmer~\citep{Lighthill:1952,Blake:1971}, which propels by a prescribed surface-slip velocity. This model has been successful in explaining the experimentally-observed behavior of the algae Volvox~\citep{Pedley:2016}, the Janus colloid~\citep{Campbell:2019}, chemically active droplets~\citep{Maass:2016} and has been theoretically used to understand the near-field hydrodynamic interaction with a flat boundary~\citep{Ishimoto:2013,Shaik:2017,Poddar:2020,Poddar:2023,Ghosh:2023,Shum:2025} and in channels~\citep{Zhu:2013}. The aspect of surface topography for this model has been addressed in Ref.~\citep{Ishimoto:2023} using the method of regularized Stokeslets for some specific parameters, highlighting that boundary guidance is possible if the surface wavelength and the squirmer size are comparable. Yet, a comprehensive understanding of active transport near structured surfaces, the dependence on the flow signature of the squirmer, and the role of hydrodynamic interactions versus steric repulsion remains elusive.

Here, we study the dynamics of a squirmer near a corrugated, periodic surface. Employing a domain-perturbation method for small surface amplitude and the Lorentz reciprocal theorem, provides semi-analytical predictions for the roughness-induced translational and angular velocities of the squirmer. Our calculations rely on a bispherical coordinate representation, thus allowing to resolve near-field flows. We exploit our semi-analytical framework, which has been benchmarked with a far-field theory and through the method of regularized Stokeslets, to explore the dependence of the squirmer dynamics on the corrugation wavelength and the initial configurations.  We reveal that pullers can become trapped in the surface valleys, while pushers tend to move along the surface corrugations. Most prominently, pullers and pushers tilted with respect to the corrugations can be sorted by varying the corrugation wavelength. 

The manuscript is organized as follows: In Section~\ref{sec:1} we present our model set-up and outline the solution framework in Section~\ref{sec:2}. In Section~\ref{sec:3} we discuss the results and we finally conclude in Section~\ref{sec:4}.

\section{Problem Setup \label{sec:1}}
	\begin{figure}[t]
		\centering
		\includegraphics[width=\textwidth]{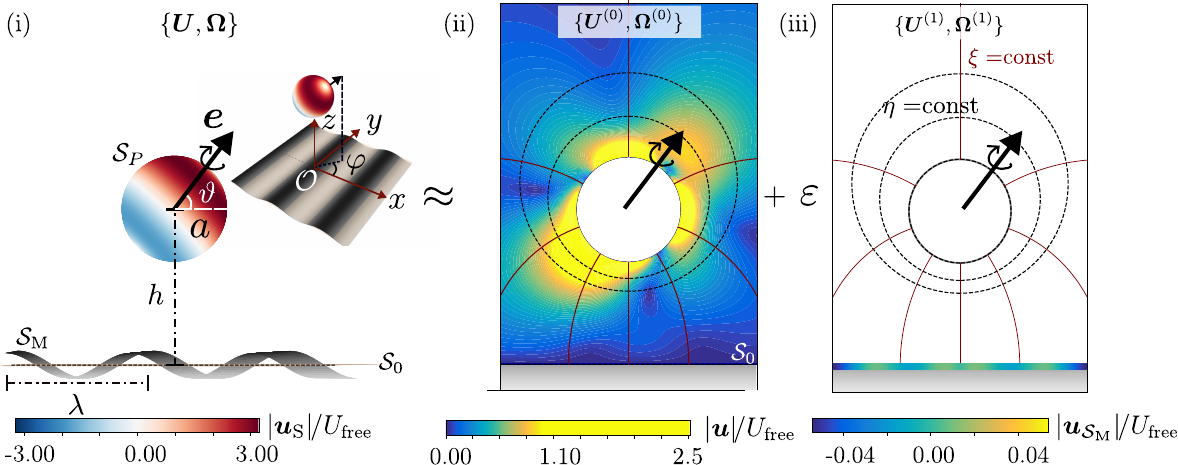}
		\caption{(i)~Model set-up of a squirmer near a periodic surface $\mathcal{S}_{\rm M}$, described by a shape function $H(x,y)$. The squirmer is oriented along $\vec{e}=(\cos\vartheta(t)\cos\varphi(t),\cos\vartheta(t)\sin\varphi(t),\sin \vartheta(t))$ and self-propels via a slip velocity~$\vec{u}_{S}$ on its surface $\mathcal{S}_{\rm P}$. It has a radius $a$ and is at a height $h$ above a planar reference surface $\mathcal{S}_{0}$ with respect to the coordinate axes $(x,y,z)$ at~$\mathcal{O}$. For small surface amplitude, the full problem is decomposed into the sum of (ii)~the squirmer near a planar no-slip wall (i.e., the zeroth-order problem) and (iii) a sphere near a wall with a slip boundary condition (i.e., the first-order perturbation of the surface corrugations). Panel (ii) shows the squirmer near a planar wall with the associated flow field. Here, the squirmer with $\beta =3$ is oriented at $\vartheta=\pi/4$ and $\varphi=0$ and at a height $h=3a$ from $\mathcal{S}_0$. The flow field is obtained using bispherical coordinates, as indicated by the lines. The constant surfaces of $\xi={\rm const.}$\,(red solid lines) and $\eta={\rm const.}$\,(black dashed lines) intersect perpendicularly. The $\phi$ coordinate is normal to $\eta$ and $\xi$, forming a right-handed coordinate system. In panel (iii) the slip velocity at the boundary $\mathcal{S}_0$ depends on the surface shape~$H(x,y)$. Here we show the slip velocity for $a\varepsilon H(x,y)=0.1\cos(2\pi x/3)$.}
		\label{fig:schematic}
	\end{figure}
We consider a squirmer of radius $a$ with tangential slip-velocity $\vec{u}_{\rm S}$ on its surface $\mathcal{S}_{\rm P}$. The latter induces self-propulsion and generates a flow with a spatially-varying velocity $\vec{u}(\vec{r})$ and pressure field $p(\vec{r})$. The swimmer moves near a structured, periodic surface $\mathcal{S}_{\rm M}$, characterized by the wavelength $\lambda$, which modifies its translational and angular velocities, $\vec{U}$ and $\vec{\Omega}$, due to hydrodynamic interactions. In our model set-up (see Fig.~\ref{fig:schematic}), the periodic surface is defined about a planar reference surface~$\mathcal{S}_{0}$, which contains the  origin $\mathcal{O}$ of our coordinate system $(x,y,z)$. Further, $\unitvec{z}$ denotes the unit normal to $\mathcal{S}_0$ and $\vec{r}_\parallel = (x,y,0)^T$ are the in-plane coordinates. 

At a given instant of time $t$, the position of the squirmer $\vec{r}_S(t)$ is at a height $h(t)$ above the reference surface $\mathcal{S}_0$. The orientation of the swimmer is defined by the vector $\vec{e}=\cos\vartheta(t)(\cos\varphi(t)\unitvec{x}+\sin \varphi(t)\unitvec{y})+\sin\vartheta(t)\unitvec{z}$, where $\vartheta(t)$ and $\varphi(t)$ are the angles with respect to $x-$ and $y-$axes, normal to and on $\mathcal{S}_0$, respectively. We consider the limit of vanishing inertial forces of the squirmer compared to the viscous forces of the fluid, i.e. $U_{\rm free}a\ll \mu/\rho$, where $U_{\rm free}$ is the typical squirmer velocity, $\mu$ and $\rho$ are the fluid viscosity and density, respectively. Thus, the velocity and pressure fields \,$(\vec{u}(\vec{r}),p(\vec{r}))$ at a position $\vec{r}$ outside the squirmer are governed by the Stokes equations:
\begin{align}
\vec{\nabla}\cdot\vec{\sigma} &= \vec{0} \quad {\rm and} \quad \vec{\nabla}\cdot\vec{u}=0, \label{eq:stokes}
\end{align}
with stress tensor $\vec{\sigma} = -p\mathbb{I}+\mu[\vec{\nabla}\vec{u}+(\vec{\nabla} \vec{u})^T]$, where and $\mathbb{I}$ denotes the unit matrix. The equations are complemented by the boundary conditions (BCs) on the surface of the swimmer~$\mathcal{S}_{\rm P}$ and the structured wall~$\mathcal{S}_{\rm M}$: 
\begin{subequations}
\begin{align}
    \vec{u}&=\vec{u}_{S}+\vec{U}+\vec{\Omega}\times (\vec{r}-\vec{r}_{S}) &\quad &{\rm on}\quad \mathcal{S}_{\rm P},\label{eq:BC_sphere}\\
    \vec{u}&=\vec{0} &\quad &{\rm on}\quad \mathcal{S}_{\rm M}. \label{eq:BC_surface}
\end{align}    
\end{subequations}
Further, the velocity vanishes far away from the swimmer, $\vec{u}=\vec{0}$ on $\mathcal{S}_\infty$. In this work, we consider a tangential slip velocity $\vec{u}_{S}$ with azimuthal symmetry. Following the work of~\citep{Shaik:2017,Poddar:2020,Shum:2025}, we focus on the first two longitudinal modes. In a coordinate-independent notation, the slip velocity assumes the form
\begin{align}\label{eq:squirmer}
\boldsymbol{u}_{S}=B_{1}(1+\beta \vec{e}\cdot\vec{r}')(\vec{r}'\vec{r}'-\mathbb{I})\cdot\vec{e}, 
\end{align}
where $\vec{r}'=(\vec{r}-\vec{r}_S)/a$ is the unit vector, pointing from the center of the squirmer to a point on its surface. Here, $B_1$ sets the swim speed in free space via $\vec{U}_{\rm free}=2B_1\vec{e}/3$. The `squirmer parameter' $\beta=B_{2}/B_{1}$ corresponds to the dipole strength in the multipole expansion of the flow field of a squirmer in free space.  Depending on its sign, the squirmer is classified as a puller\,($\beta>0$), which pulls the fluid from the front and the back and pushes it from the sides, a pusher \,($\beta<0$), which pushes the fluid from the front and the back and pulls it in from the side, and the neutral swimmer\,($\beta=0$),  having a zero dipolar flow field.

In the absence of inertia or external forces and torques, the  hydrodynamic force $\vec{F}_{\rm H}$ and torque $\vec{L}_{\rm H}$ experienced by the squirmer vanish:
\begin{align}\label{eq:force_torque_free_condition}
    \vec{F}_{\rm H}=\int_{\mathcal{S}_{\rm P}} \vec{n}\cdot\vec{\sigma}\, \diff S=\vec{0} \quad {\rm and}\quad  \vec{L}_{\rm H}=\int_{\mathcal{S}_{\rm P}}  (\vec{r}-\vec{r}_{S})\times (\vec{n}\cdot\vec{\sigma})\, \diff S=\vec{0}, 
\end{align}
where $\vec{n}$ is the unit normal to the surface of the squirmer. Eq.~\eqref{eq:force_torque_free_condition} establishes a relation between the prescribed slip velocity $\vec{u}_{\rm S}$ and the swimming velocities $(\vec{U},\vec{\Omega})$. 

Lastly, we model the periodic surface $\mathcal{S}_{\rm M}$ in the Monge-gauge representation through the surface function $F\equiv z-\varepsilon aH(\vec{r}_{\parallel})$ with surface shape
\begin{align}\label{eq:surface_function}
        H(\vec{r}_{\parallel})=\cos \left(\frac{2\pi x}{\lambda}\right).
\end{align}
Here,  $\varepsilon a$ is the amplitude of the surface and $\lambda$ denotes the surface wavelength. While for the purpose of this work we consider a cosine surface, it is worth mentioning that any smooth surface can be represented as sum of cosines and sines, and thus this approach is readily extendable to arbitrary surface shapes~\citep{Kurzthaler:2020}. 

\subsection{Short-ranged repulsive force}
Close to the surface the squirmer experiences a steric interaction of non-hydrodynamic origin that may affect its swimming dynamics. Following the work of~\citep{Brady:1985}, we model the steric interaction with the structured surface through a short-ranged repulsive force of the form 
\begin{align}
    \vec{F}_{\rm rep}=\frac{A\exp\left\{-B\left(\delta-a\right)\right\}}{1-\exp\left\{-B\left(\delta-a\right)\right\}}\vec{\delta}\equiv F_{\rm rep}(\delta)\vec{\delta}, \label{eq:repulsive_force}
\end{align}
where the parameters $A$ and $B$ set the strength and the range of the repulsive force, respectively.  Further, we introduced the unit vector $\vec{\delta}=[\vec{r}_{S}-\vec{r}_{\mathcal{S}_{\rm M}}]_{\rm min}/\delta$, where $\delta=\norm{\vec{r}_{S}-\vec{r}_{\mathcal{S}_{\rm M}}}_{\rm min}$ is the minimum distance of the squirmer center ($\vec{r}_S=(x_S, y_S, h)$) to the surface (minimized over the set of all points $\vec{r}_{\mathcal{S}_{\rm M}}=(x,y,\varepsilon a H(x,y))\in\mathcal{S}_M$).
Let the point on the surface that minimizes the distance be referred to as $\vec{r}_m=\left(x_m,y_m,\varepsilon a H(x_m,y_m)\right)$. We obtain this point by minimizing $\delta$ and explicitly express the distance from the center of the sphere to an arbitrary point on the surface $\vec{r}_{\mathcal{S}_{\rm M}}$, through
\begin{align}
    \delta_{\mathcal{S}_{\rm M}}=\sqrt{\left(x_S-x\right)^2+\left(y_S-y\right)^2+\left(h-\varepsilon a H\left(x,y\right)\right)^2}.
\end{align}
The closest point $\vec{r}_m$ is obtained by using the minimization condition,
\begin{align}
    \frac{\partial \delta_{\mathcal{S}_{\rm M}}}{\partial x}=0\quad {\rm and}\quad   \frac{\partial \delta_{\mathcal{S}_{\rm M}}}{\partial y}=0\quad {\rm at}\quad \vec{r}_{\mathcal{S}_M}=\vec{r}_m.
\end{align}
Evaluating the derivative and simplifying, two general expressions result:
\begin{align}
    x_{m}&=x_S+\left(h-\varepsilon a H\left(x_m,y_m\right)\right)\varepsilon a \,\left.\frac{\partial H(x,y)}{\partial x}\right|_{(x_m,y_m)}\quad {\rm and}\label{eq:x_m}\\
    y_{m}&=y_S+\left(h-\varepsilon a H\left(x_m,y_m\right)\right)\varepsilon a \,\left.\frac{\partial H(x,y)}{\partial y}\right|_{(x_m,y_m)}.\label{eq:y_m}
\end{align}
Having evaluated $\delta=|\vec{r}_{S}-\vec{r}_{m}|$, we calculate the unit-direction $\vec{\delta}$ using its definition
\begin{align}
    \vec{\delta}=\frac{(x_S-x_m)\unitvec{x}+(y_S-y_m)\unitvec{y}+(h-\varepsilon a H(x_m,y_m))\unitvec{z}}{\sqrt{(x_S-x_m)^2+(y_S-y_m)^2+\left(h-\varepsilon a H(x_m,y_m)\right)^2}}.
\end{align}
Substituting for $x_S-x_m$ and $y_S-y_m$ from Eqs.~\eqref{eq:x_m} and~\eqref{eq:y_m}, respectively, we note that
\begin{align}
\vec{\delta}=\frac{-\varepsilon a \left[\grad H(x,y)\right]_{(x_m,y_m)}+\unitvec{z}}{\sqrt{ \varepsilon^2 a^2 \left|\left[\grad H(x,y)\right]_{(x_m,y_m)}\right|^2+1}}=\frac{\grad F}{|\grad F|}\equiv\vec{n},
\end{align}
 at $\vec{r}_m$ on $\mathcal{S}_{\rm M}$. Thus, the shortest distance from the center of the sphere to the surface aligns with the surface normal. As this force acts along the center of the squirmer, it results only in a repulsive force and no torque. Thus, introducing a short-range repulsive force, modifies the force balance via: 
\begin{align}
\vec{F}_{\rm H}+ \vec{F}_{\rm rep}=\vec{0},
\end{align}
while keeping the torque-free condition [Eq.~\eqref{eq:force_torque_free_condition}] unchanged. 

\section{Solution methodology \label{sec:2}} 
To make analytical progress, we follow a perturbation approach in the surface amplitude.  We first perform a domain perturbation and expand the set of equations and the repulsive force in the perturbation parameter. Employing the Lorentz reciprocal theorem, we derive the swimming velocities, which we compute using a bispherical coordinate representation. 

\subsection{Domain-perturbation method}
We solve the problem in the regime where the surface amplitude is small compared to the size of the particle, corresponding to $\varepsilon\ll1$. Therefore, we first expand the velocity boundary condition on the surface $\mathcal{S}_{\rm M}$ (corresponding to $z=\varepsilon a H(x,y)$) in a Taylor series about the reference surface $\mathcal{S}_{0}$ (corresponding to $z=0$):
\begin{align}\label{eq:BC_taylor}
    \vec{u}(z=\varepsilon aH(x,y))=\vec{u}(z=0)+\varepsilon a H(x,y)\left.\frac{\partial \vec{u}}{\partial z}\right|_{z=0}+\mathcal{O}(\varepsilon^{2}).
\end{align} 
Furthermore, we expand all our solution variables in the form
\begin{align}\label{eq:perturbation}
\begin{split}
    \{\vec{u},p,\vec{\sigma},\vec{U},\vec{\Omega}\}= \{\vec{u}^{(0)},p^{(0)},&\vec{\sigma}^{(0)},\vec{U}^{(0)},\vec{\Omega}^{(0)}\} \\
    &+ \varepsilon\{\vec{u}^{(1)},p^{(1)},\vec{\sigma}^{(1)},\vec{U}^{(1)},\vec{\Omega}^{(1)}\}+\mathcal{O}(\varepsilon^{2}).
\end{split}
\end{align}
The perturbation expansion decomposes the full problem into a recursive scheme (see Fig.~\ref{fig:schematic} for a schematic): Substituting the perturbation expansions [Eqs.~\eqref{eq:BC_taylor}-\eqref{eq:perturbation}] into the Stokes equations [Eqs.~\eqref{eq:stokes}] and the boundary conditions [Eqs.~\eqref{eq:BC_sphere}-\eqref{eq:BC_surface}], we obtain at zeroth-order the problem of a squirmer near a rigid surface:
\begin{subequations}
    \begin{align}
      \vec{\nabla}\cdot \vec{\sigma}^{(0)}&=\vec{0}\quad {\rm and} \quad \vec{\nabla}\cdot \vec{u}^{(0)}=0, \label{eq:zeroth}\\
       \vec{u}^{(0)}&=\boldsymbol{0} &\quad &{\rm on}\quad \mathcal{S}_{\rm \infty},\ \mathcal{S}_{0}, \label{eq:zeroth_bc_1}\\
    \vec{u}^{(0)}&=\vec{u}_{S}+\vec{U}^{(0)}+\vec{\Omega}^{(0)}\times \vec{r}' &\quad &{\rm on}\quad \mathcal{S}_{\rm P}.\label{eq:zeroth_bc_2}
    \end{align}
\end{subequations}
This set of equations can be solved semi-analytically using a bispherical coordinate approach~\citep{Poddar:2020} (see Sec.~\ref{sec:bispherical} and Appendix~\ref{appA}). The first-order problem depends on the zeroth-order solution via:
\begin{subequations}
    \begin{align}
      \vec{\nabla}\cdot\vec{\sigma}^{(1)}&=\vec{0}\quad {\rm and} \quad \vec{\nabla}\cdot \vec{u}^{(1)}=0\label{eq:first}\\
      \vec{u}^{(1)}&=\boldsymbol{0}& &{\rm on}\quad \mathcal{S}_{\infty},\\ 
      \vec{u}^{(1)}&=-aH(\vec{r}_\parallel)\frac{\partial \vec{u}^{(0)}}{\partial z}\equiv \vec{u}_{\mathcal{S}_M}& &{\rm on}\quad \mathcal{S}_{0},  \label{eq:first_bc_surface} \\
      \vec{u}^{(1)}&=\vec{U}^{(1)}+\vec{\Omega}^{(1)}\times \vec{r}' &  &{\rm on}\quad \mathcal{S}_{\rm P}.\label{eq:first_bc_sphere}
    \end{align}
\end{subequations}
This problem represents a sphere translating and rotating near a surface with a slip velocity~$\vec{u}_{\mathcal{S}_M}$, which depends on the surface shape $H(\vec{r}_\parallel)$ and the zeroth-order velocity field. A central aim of this work is to calculate the first-order swimming velocities ($\vec{U}^{(1)}$ and $\vec{\Omega}^{(1)}$) and use them as input for studying the leading-order boundary-induced correction to the squirming dynamics. 

\subsection{Small $\varepsilon$-expansion of the repulsive force} 
As last step of our perturbation approach, we expand the repulsive force [Eq.~\eqref{eq:repulsive_force}] in $\varepsilon$. Substituting for the surface function through Eq.~\eqref{eq:surface_function} and expanding Eqs.~\eqref{eq:x_m} and ~\eqref{eq:y_m}
in $\varepsilon$, we find a transcendental equation for $x_m$
 \begin{align}
x_{{\rm m}}+\varepsilon h\frac{2\pi a}{\lambda}\sin\left(\frac{2\pi x_{\rm m}}{\lambda}\right)-x_S+\mathcal{O}(\varepsilon^2)=0, \label{eq:transcendental}
\end{align}
which needs to be solved numerically, while we recover $y_m=y_S$. Expanding the repulsive force [Eq.~\eqref{eq:repulsive_force}] around $h$ in $\varepsilon$ and collecting terms up to $\mathcal{O}(\varepsilon)$, leads to:  
  \begin{align}
    \vec{F}_{\rm rep}(\delta) &= F_{\rm rep}(h)(\unitvec{z}-\varepsilon a\partial_{x}H)+\left(\delta-h\right)\left.\frac{\partial F_{\rm rep}}{\partial \delta}\right|_{\delta=h}\unitvec{z}+\mathcal{O}(\varepsilon^2)\nonumber\\
    &=F_{\rm rep}(h)\unitvec{z}+\left(-\varepsilon a F_{\rm rep}(\delta)\frac{\partial H}{\partial x}\,\unitvec{x}+\left(\delta-h\right)\left.\frac{\partial F}{\partial \delta}\right|_{\delta=h}\unitvec{z}\right)\nonumber\\
    &=\vec{{F}}_{\rm rep}^{(0)}+\varepsilon\vec{{F}}_{\rm rep}^{(1)}\nonumber
\end{align}
where we identified the zeroth- and first-order contributions: 
\begin{subequations}
\begin{align}
    \vec{{F}}_{\rm rep}^{(0)}&=\frac{A\exp\left\{-B\left(h-a\right)\right\}}{1-\exp\left\{-B\left(h-a\right)\right\}}\unitvec{z} \quad{\rm and},\label{eq:repulsive_zero}\\ \vec{{F}}_{\rm rep}^{(1)}&=\frac{A\exp\left\{-B\left(h-a\right)\right\}}{1-\exp\left\{-B\left(h-a\right)\right\}}\left[\frac{2\pi a}{\lambda}\sin\left(\frac{2\pi 
x_{\rm m}}{\lambda}\right)\unitvec{x}-\frac{B(\delta-h)}{\varepsilon\left(1-\exp\left\{-B\left(h-a\right)\right\}\right)}\unitvec{z}\right]\label{eq:repulsive_first}.
\end{align} 
\end{subequations}
Here, it is important to note that $\delta-h\sim \mathcal{O}(\varepsilon)$. We can show this in the limit of large $\lambda$ explicitly, by approximating $x_m$ as
\begin{align}
 x_{{\rm m}}\approx x_{S}-\varepsilon  h a \left(\frac{2\pi}{\lambda}\right)^2 x_S.\label{eq:xm} 
\end{align}
Substituting Eq.~\eqref{eq:xm} into $\delta$, we obtain to leading order in $\varepsilon$:
\begin{align}
    \delta&
    \approx h\left[1-\varepsilon\frac{a}{h}  H(x_{{\rm m}},y_{{\rm m}})\right],\label{eq:dm}
\end{align}
hence showing that $\delta-h\sim \mathcal{O}(\varepsilon)$. 

Further, the force at zeroth order [Eq.~\eqref{eq:repulsive_zero}] recovers the force typically used for the planar wall case~\citep{Poddar:2020, Garai:2025}, while the first-order force contains a contribution along the $\unitvec{z}$ and $\unitvec{x}$ directions. The repulsive force then enters the zeroth- and first-order problem via modifying the force balances as
\begin{align}
\vec{{F}}^{(0)}_{\rm H}+\vec{{F}}^{(0)}_{\rm rep}=\vec{0}  \quad{\rm and}\quad \vec{{F}}^{(1)}_{\rm H}+\vec{{F}}^{(1)}_{\rm rep}=\vec{0}.\label{eq:squirmer_force_balance_perturbation}
\end{align}

A short note on the definition of $\delta$ which we use in practice is in order here. We measure the distance $\delta$ from $z=\varepsilon a$ instead of $z=0$. This is a numerical simplification to avoid spurious values when the squirmer is on top of a `valley'  of the underlying surface, and the minimum distance between the swimmer and the surface $\delta<\varepsilon$. In essence this can be controlled by varying the parameter $B$ in Eq.~\eqref{eq:repulsive_force}, but doing so increases the range of the repulsive force. While this would not lead to a significant change in the swimmer dynamics (see Appendix~\ref{appC}), it would increase the minimum distance of approach $h_{\rm min}$. Thus, to limit the study to a short-ranged force, the swimmer-surface distance is measured from $z=\varepsilon a$ (instead of $z=0$).
\subsection{Reciprocal theorem} 
Our aim is to calculate the leading-order swimming velocities $\vec{U}=\vec{U}^{(0)}+\varepsilon\vec{U}^{(1)}$ and $\vec{\Omega}=\vec{\Omega}^{(0)}+\varepsilon\vec{\Omega}^{(1)}$. This amounts to solving Eqs.~\eqref{eq:zeroth}-~\eqref{eq:zeroth_bc_2} and Eqs.~\eqref{eq:first}-~\eqref{eq:first_bc_sphere}, given the slip velocity of the squirmer $\vec{u}_{S}$ [Eq.~\eqref{eq:squirmer}], and using the modified force balance [Eq.~\eqref{eq:squirmer_force_balance_perturbation}] and torque-free condition  [Eq.~\eqref{eq:force_torque_free_condition}]. A convenient way to calculate them is to utilize the Lorentz reciprocal theorem for viscous flows~\citep{Masoud:2019}. It states that two problems $\left(\vec{u}, \vec{\sigma}\right)$ and  $\left(\vec{\hat{u}},\vec{\hat{\sigma}}\right)$, which share the same geometry with boundaries $\mathcal{S}$ but are characterized by different boundary conditions on $\mathcal{S}$, are related through the integral relation:
\begin{align}\label{eq:reciprocal}
     \int_{\mathcal{S}} \vec{n}\cdot \vec{\hat{\sigma}}\cdot \vec{u} \,\diff S=     \int_{\mathcal{S}} \vec{n}\cdot \vec{\sigma}\cdot \vec{\hat{u}} \,\diff S,
\end{align} 
where $\vec{n}$ is the unit normal pointing from the surface $\mathcal{S}$ into the fluid. For our problem, we can exploit that the problem of a sphere translating and rotating at velocities $\vec{\hat{U}}$ and $\vec{\hat{\Omega}}$ under externally-applied forces $\vec{\hat{F}}$ and torques $\vec{\hat{L}}$ near a planar, no-slip surface has been extensively studied~\citep{Stimson:1926,Brenner:1961,Dean:1963,O’Neill:1964,Lee:1980}. The boundary conditions associated to this auxiliary problem are: $\vec{\hat{u}}=\vec{\hat{U}}+\vec{\hat{\Omega}}\times\vec{r}'$ on $\mathcal{S}_{\rm P}$, and $\vec{\hat{u}}=\vec{0}$ on~$\mathcal{S}_0$ and $\mathcal{S}_\infty$; and the force and torque balance read: $\vec{\hat{F}}+\vec{\hat{F}}_H=\vec{0}$ and $\vec{\hat{L}}+\vec{\hat{L}}_H=\vec{0}$. 

We next substitute the corresponding solution $\{\vec{\hat{u}},\vec{\hat{\sigma}}\}$ and our swimmer problem $\{\vec{u}=\vec{u}^{(0)}+\varepsilon \vec{u}^{(1)}, \vec{\sigma}=\vec{\sigma}^{(0)}+\varepsilon \vec{\sigma}^{(1)} \}$ into Eq.~\eqref{eq:reciprocal}. For convenience, we introduce the generalized swimming velocities and the hydrodynamic forces $\vec{\mathcal{U}}\equiv (\vec{U},\vec{\Omega})^{T}$ and   $\vec{\mathcal{F}}_{\rm H}\equiv (\vec{F}_{\rm H},\vec{L}_{\rm H})^{T}$, respectively. This simplifies the reciprocal relation [Eq.~\eqref{eq:reciprocal}] to
\begin{align}\label{eq:reciprocal_2}
\begin{split}
\vec{\mathcal{U}}^{(0)}\cdot\vec{\mathcal{\hat{F}}}_{\rm H}+\int_{\mathcal{S}_{P}} \vec{n}\cdot \vec{\hat{\sigma}}\,\cdot \,&\vec{u}_{\rm S} \,\diff S +\varepsilon \vec{\mathcal{U}}^{(1)}\cdot\vec{\mathcal{\hat{F}}}_{\rm H}\\
&+\varepsilon\int_{\mathcal{S}_{0}} \vec{n}\cdot \vec{\hat{\sigma}}\cdot\, \vec{u}_{\mathcal{S}_{\rm M}}\, \diff S= \vec{\mathcal{\hat{U}}}\cdot\left(\vec{\mathcal{F}}^{(0)}_{\rm H}+\varepsilon \vec{\mathcal{F}}^{(1)}_{\rm H}\right)
\end{split}
\end{align}
Typically, the right-hand-side vanishes for a force-free swimmer. Here, however, this term is replaced by the short-range repulsive force [Eq.~\eqref{eq:squirmer_force_balance_perturbation}]. To simplify the expression, we express the hydrodynamic force and the stress tensor for the sphere in the auxiliary problem via $\vec{\mathcal{\hat{F}}}_{\rm H}=-\mat{R}\cdot \mat{U}$ and $\vec{\hat{\sigma}}=\mat{T}\cdot \mat{U}$, where $\mat{R}$ denotes the generalized resistance matrix. Thus, the total swimming velocity of the squirmer up to first order in $\varepsilon$ results in:
\begin{align}\label{eq:vel_correction_repulsive}
\vec{\mathcal{U}}=\mat{R}^{-1}\cdot\left[\int_{\mathcal{S}_{P}} \vec{n}\cdot\vec{\mathcal{\hat{T}}}\cdot \vec{u}_{\rm S}\,\diff S+\vec{\mathcal{F}}^{(0)}_{\rm rep}+\varepsilon\left( \int_{\mathcal{S}_{0}}\vec{n}\cdot \mat{T}\cdot \vec{u}_{\mathcal{S}_{\rm M}}\,\diff S+\vec{\mathcal{F}}^{(1)}_{\rm rep}\right)\right]+\mathcal{O}(\varepsilon^{2}).
\end{align}
In the next section, we will show how this expression can be evaluated numerically using a bispherical coordinate expansion of the stress and velocity fields.

\subsection{Bispherical coordinate representation \label{sec:bispherical}}
Taking advantage of solving the Stokes equation in a domain of a sphere near a flat surface\,(corresponding to a sphere of infinite radius), we revert to a bispherical coordinate representation ($\xi$, $\eta$, $\phi$) of our velocity and pressure fields. This enables solving for the corrugation-induced velocities [Eq.~\eqref{eq:vel_correction_repulsive}] semi-analytically. In what follows we omit specifying between the auxiliary ($\vec{\hat{u}}$, $\hat{p}$) and the main (zeroth-order) problem ($\vec{{u}}^{(0)}$, $p^{(0)}$), as the general representation in bispherical coordinates remains the same.
The transformation from the cartesian coordinates reads
\begin{align}
    x=\frac{c\sin \xi \cos\phi}{\cosh \eta-\cos\xi},\quad  y=\frac{c\sin \xi \sin\phi}{\cosh \eta-\cos\xi},\quad z=\frac{c\sinh \eta}{\cosh \eta-\cos\xi},
\end{align}
where $c=a{\rm sinh}(\eta_0)$ is set by the geometry with $\eta_0 = {\rm arccosh}(h/a)$. Note that these are related to cylindrical coordinates $(\rho, \phi, z)$ via $\rho=\sqrt{x^2+y^2}$ and $\phi=\arctan(y/x)$. A general consequence of the linearity of the incompressible Stokes equations [Eq.~\eqref{eq:stokes}] is that the pressure is harmonic $\nabla^2p=0$. In our geometrical setup, the Laplace equation is separable in bispherical coordinates, allowing for a solution of the form~\citep{Lee:1980}
\begin{subequations}
\begin{align}
p&=\sum_{m=0}^{\infty}p_{m}\cos(m\phi+\alpha_{m}),\label{eq:pressure}
\end{align}
with coefficients
\begin{align}
p_m &= \frac{1}{c}(\cosh \eta-\mu)^{\frac{1}{2}}\sum_{n=m}^{\infty}\left[A_{n}^{m}\sinh \left(n+\frac{1}{2}\right)\eta+B_{n}^{m}\cosh \left(n+\frac{1}{2}\right)\eta\right]P_{n}^{m}(\mu), 
\end{align}
\end{subequations} 
where $P_n^m(\cdot)$ are the associated Legendre polynomials of the first kind of degree $n$ and order $m$ and $\alpha_m$ is a phase for the $m^{\rm th}-$mode, which depends on the problem. The components of the velocity field in a cylindrical coordinate system\,($\boldsymbol{u}=u\unitvec{\rho}+v\unitvec{\phi}+w\boldsymbol{\hat{z}}$) are expressed as sum of a particular and a homogeneous solution:
\begin{subequations}
\begin{align}
u&=\frac{\rho p}{2}+u_{0}\cos \alpha_{0}+\frac{1}{2}\sum_{m=1}^{\infty}(\gamma_{m}+\chi_{m})\cos(m\phi+\alpha_{m}),\label{eq:u}\\
v&=v_{0}\sin \alpha_{0}+\frac{1}{2}\sum_{m=1}^{\infty}(\gamma_{m}-\chi_{m})\sin(m\phi+\alpha_{m}),\,\,{\rm and}\label{eq:v}\\
w&=\frac{zp}{2}+\sum_{m=0}^{\infty}w_{m}\cos(m\phi+\alpha_{m})\label{eq:w},
\end{align}
where the coefficients are 
	\begin{align}
u_{0}&=(\cosh \eta-\mu)^{\frac{1}{2}}\sum_{n=1}^{\infty}\left[E_{n}^{0}\sinh \left(n+\frac{1}{2}\right)\eta+F_{n}^{0}\cosh \left(n+\frac{1}{2}\right)\eta\right]P_{n}^{1}(\mu),\label{eq:u0}\\
v_{0}&=(\cosh \eta-\mu)^{\frac{1}{2}}\sum_{n=1}^{\infty}\left[G_{n}^{0}\sinh \left(n+\frac{1}{2}\right)\eta+H_{n}^{0}\cosh \left(n+\frac{1}{2}\right)\eta\right]P_{n}^{1}(\mu),\label{eq:v0}\\
\gamma_{m}&=(\cosh \eta-\mu)^{\frac{1}{2}}\sum_{n=m+1}^{\infty}\left[E_{n}^{m}\sinh \left(n+\frac{1}{2}\right)\eta+F_{n}^{m}\cosh \left(n+\frac{1}{2}\right)\eta\right]P_{n}^{m+1}(\mu),\label{eq:gamma_m}\\
\chi_{m}&=(\cosh \eta-\mu)^{\frac{1}{2}}\sum_{n=m-1}^{\infty}\left[G_{n}^{m}\sinh \left(n+\frac{1}{2}\right)\eta+H_{n}^{m}\cosh \left(n+\frac{1}{2}\right)\eta\right]P_{n}^{m-1}(\mu),\,\label{eq:chi_m}\\
w_{m}&=(\cosh \eta-\mu)^{\frac{1}{2}}\sum_{n=m}^{\infty}\left[C_{n}^{m}\sinh \left(n+\frac{1}{2}\right)\eta+D_{n}^{m}\cosh \left(n+\frac{1}{2}\right)\eta\right]P_{n}^{m}(\mu).\label{eq:w_m}
\end{align}
\end{subequations} 
The relation between the eight coefficients\,$\{A_{n}^{m},B_{n}^{m},C_{n}^{m},D_{n}^{m},E_{n}^{m},F_{n}^{m},G_{n}^{m},H_{n}^{m}\}$ are obtained using the incompressibility condition and the respective boundary conditions of the main zeroth-order \,[Eqs.~\eqref{eq:zeroth_bc_1}-~~\eqref{eq:zeroth_bc_2}] or the auxiliary problem. These are projected in the bispherical basis, leading to relations between the coefficients listed in the Appendix~\eqref{appA}. 

It is worth pointing out that the solution to the Stokes equation in bispherical coordinates in Eqs.~\eqref{eq:pressure} and \eqref{eq:u}-~\eqref{eq:w} are represented in terms of an angular decomposition\,(in $\phi$) in different orders of $m$. A similar decomposition is obtained for the squirming velocity\,[Eq.~\eqref{eq:squirmer}] by projecting it on the bispherical coordinates (see Appendix~\eqref{squirmerbc}). The zeroth-order swimming velocity is then readily evaluated by exploiting the orthogonality relations of the angular modes in the first integral of Eq.~\eqref{eq:vel_correction_repulsive}. A similar simplification is obtained for the first-order swimming velocity by decomposing the effective slip velocity $\vec{u}_{\mathcal{S}_{\rm M}}$ in angular modes in the second integral of Eq.~\eqref{eq:vel_correction_repulsive}. To this end, we express the surface shape function in terms of the Jacobi-Anger expansion~\citep{Abramowitz:1964}
\begin{align}\label{eq:jacobi}
 H(\rho,\phi;x_{S})&=\cos \left(\frac{2\pi \rho \cos\phi}{\lambda}\right)\cos\left(\frac{2\pi x_{S}}{\lambda}\right)-\sin \left(\frac{2\pi \rho \cos\phi}{\lambda}\right)\sin\left(\frac{2\pi x_{S}}{\lambda}\right)\nonumber\\&\equiv \cos\left(\frac{2\pi x_{S}}{\lambda}\right)\left\{J_{0}\left(\frac{2\pi\rho}{\lambda}\right)+2\sum_{m=1}^{\infty}(-1)^{m}J_{2m}\left(\frac{2\pi\rho}{\lambda}\right)\cos(2m\phi)\right\}\nonumber\\
 &-\sin \left(\frac{2\pi x_{S}}{\lambda}\right)\left\{2\sum_{m=0}^{\infty}(-1)^{m}J_{2m+1}\left(\frac{2\pi\rho}{\lambda}\right)\cos((2m+1)\phi)\right\},
\end{align}
where $\rho,\phi$ are the cylindrical coordinates with respect to $(x_S, y_S, 0)$ and $J_m(\cdot)$ is the Bessel function of the first kind of order $m$. Note that Eq.~\eqref{eq:jacobi} can be readily generalized for any surface function as long as it can be expressed in Fourier modes in two dimensions. Through our framework, the squirmer velocities near any such surface can then be obtained through evaluation of Eq.~\eqref{eq:vel_correction_repulsive}.

\subsection{Numerical validation}
We validate our perturbation results with the Method of Regularized Stokeslet Surfaces\,(MRS)\,\citep{Ferranti:2024}\,(see Appendix\,~\eqref{app:validation}). We also use the latter to compute selected velocity fields, allowing us to rationalize some of our findings. To do so, we solve the Stokes equations for a force- and torque-free spherical squirmer in free space with a prescribed slip velocity at the surface. Following \citep{Ferranti:2024}, we discretize the spherical surface with a triangular mesh consisting of 320 triangles and $N_\text{sphere}=162$ vertices above a wall. To approximate an infinite corrugated surface, we discretize a $12a \times 12a$ square using another triangular mesh consisting of 5,000 triangles and $N_\text{surface}=2,601$ vertices and velocity uniformly equal to zero. We prescribe the regularization parameter $\epsilon_\text{reg}=10^{-4}$. This system of $3(N_\text{sphere}+N_\text{surface})$ equations together with 6 equations for the force and torque balance, allows solving for the $3(N_\text{sphere}+N_\text{surface})+6$ unknown velocity components at each vertex, yielding $\vec{U}$ and $\vec{\Omega}$.

\subsection{Kinematics}
To study swimming dynamics of squirmers near a corrugated surface, we evolve the swimmer position $\vec{r}_{\rm S}(t)$ and its orientation $\vec{e}(t)$ over time $t$ by numerically solving 
\begin{align}
    \frac{\diff \vec{r}_{\rm S}}{\diff t}=\vec{U}\quad {\rm and} \quad  \frac{\diff \vec{e}}{\diff t}=\vec{\Omega}\times \vec{e},\label{eq:trajectories}
\end{align}
where the swimming velocities calculated up to the leading order in $\varepsilon$ [Eq.~\eqref{eq:vel_correction_repulsive}] are used as input. We measure length scales in units of the squirmer radius $a$ and time scales in units of $\tau = a/U_{\rm free}$ with $U_{\rm free} = 2B_1/3$ denoting the swim speed in free space. The repulsive force parameters\,($A/(\mu U_{\rm free} a)=10^7,\,B/a=90$) in Eq.~\eqref{eq:repulsive_force} have been chosen such that the closest squirmer-surface distance is $h^{\star}\approx 1.15 a$, which is close enough to observe the near-field effects while keeping the force short-ranged so that its effect is negligible for $h>1.2a$. Equations~\eqref{eq:trajectories} are evolved using the ${\rm RK}45$ integrator from the $``{\rm scipy}.{\rm intgrate}.{\rm solve}\_{\rm ivp}"$ library of Python for $t_{\rm final}=100 \tau$. The maximum time step is reduced from $\Delta t_{\rm max}=0.01\tau$ to $\Delta t_{\rm max}=0.001\tau$ when the squirmer is close to the surface ($h(t)\leq1.17a$) to reduce the numerical errors associated with the large change of the perpendicular velocity close to the surface. 

\subsection{Classification of trajectories}
We classify the long-time behavior of the squirmer by its final distance to the surface $h(t_{\rm final})$ and orientation $\vartheta(t_{\rm final})$.  In the long-time limit, the swimmer either reaches a steady position and orientation $(h(t_{\rm final}),\vartheta(t_{\rm final})=(h^\star,\vartheta^\star))$, is far away from the surface\,($h(t_{\rm final})>3h(0)$), or displays a limit-cycle behavior in $h(t)-\vartheta(t)$-space. 
For squirmers reaching a steady state of $(h(t_{\rm final}),\vartheta(t_{\rm final})=(h^\star,\vartheta^\star))$, we distinguish between spatially-trapped and sliding states, where the squirmer moves along the surface. For $\vartheta^{\star}=-0.5\pi$ ($\vartheta^{\star}=0.5\pi$) the squirmer is `\textit{trapped looking down}' (`\textit{trapped looking up}'), while for $-0.5\pi<\vartheta^{\star}<0$ ($0\leq\vartheta^{\star}<0.5\pi$) the squirmer is `\textit{sliding looking down}' (`\textit{sliding looking up}').  Swimmers that are far away from the surface, i.e., $h(t_{\rm final})\gg h(0)$ are classified as `\textit{scattering}'. Lastly, we observe `\textit{oscillating}' trajectories, characterized at long times by a limit cycle in the $h-\vartheta$ space where $h(t)$ and $\vartheta(t)$ oscillate. We identify such configurations using the Poincar\'e method in $h-\vartheta$ space, where we draw a line\,(parallel to the $h-$axis) such that it intersects the limit cycle, and quantify the number of crossings of the trajectory data across the line. When the crossings occur repeatedly at long times, we consider the final behavior to be `oscillating'. Moreover the Fast Fourier transform of the tail of the trajectory data and a visual inspection of the plot reveals the wavelength of the oscillation. 

\section{Results\label{sec:3}}
We study the leading-order effect of the periodic surface on the squirmer trajectory. For the sake of completion, we first summarize the behavior of the squirmer near the planar wall and compare it with previous works. Readers familiar with this system can directly move to Sec.~\eqref{sec:32}.

\subsection{Squirming motion near a planar wall}  
\begin{figure}[tp]
		\centering
		\includegraphics[width=\textwidth]{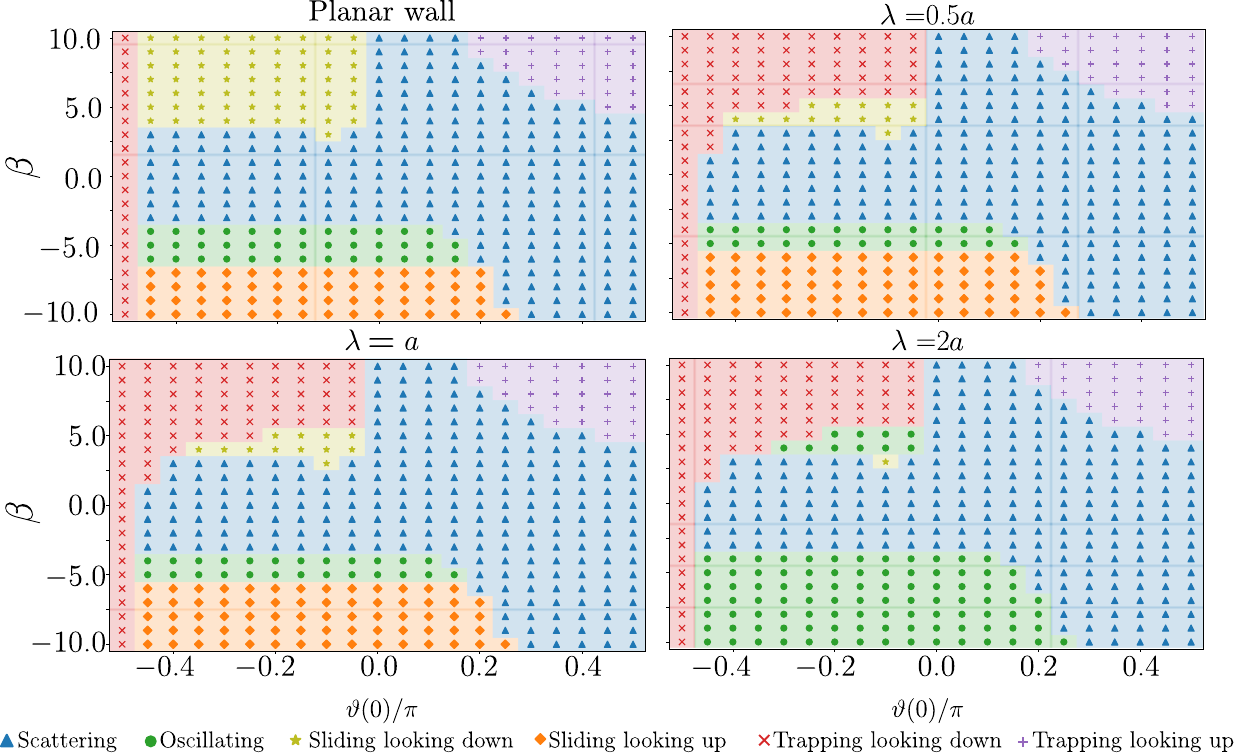}
		\caption{ Phase diagram in the $\beta-\vartheta(0)$ space. Different colors and symbols  classify different squirmer dynamics. Panels correspond to the planar wall and different wavelengths $\lambda=0.5a, a,2a$, respectively. Here, we set the initial distance to $\left(x(0),y(0),h(0)\right)=(0,0,1.5a),\,\,\varphi(0)=0,\,\,\varepsilon=0.1$, and use $A/(\mu U_{\rm free} a)=10^7$ and $B=90 a$ for the repulsive force.}
		\label{fig:phase_plot}
	\end{figure}

The behavior of a squirmer near a planar wall is governed by the details of the hydrodynamic interactions, quantified in terms of the squirmer parameter $\beta$, and the initial orientation~$\vartheta(0)$. We observe two common behaviors across all squirmers (Fig.~\eqref{fig:phase_plot}\,(i)). First, swimmers, initially oriented normal to the surface ($\vartheta(0)=-\pi/2$), are trapped looking down. Second, weak pullers\, neutral squirmers, and weak pushers ($|\beta|\leq3$) are scattered by the surface for all orientations, see Fig.~\eqref{fig:pullers_phase_portraits}\,(i) for an exemplary trajectory. Differences in behaviors manifest for larger squirmer parameters. 

Strong pullers\,($\beta>4$), initially pointing towards the surface ($\vartheta(0)<0$), end up sliding near the surface looking down (Fig.~\eqref{fig:pullers_phase_portraits}\,(ii)). Our results also show that the steady orientation decreases with the squirmer parameter $|\beta|$. While it is well known that  within the far-field description pullers have a stable configuration at $\vartheta^\star=-\pi/2$\,\citep{Spagnolie:2012}, the near-field interactions of the squirmer modify the steady orientation, leading to sliding. This is in agreement with previous work~\citep{Ishimoto:2013,Lintuvuori:2016}.  Interestingly, this behavior is reached through an initial decaying oscillatory phase, where the interplay of hydrodynamic attraction, steric-repulsion, and reorientation governs the dynamics, see trajectory in the $h-\vartheta$ space in Fig.~\eqref{fig:pullers_phase_portraits}\,(v). Lastly, strong pullers, initially pointing away from the surface,  experience strong hydrodynamic attraction and become trapped looking up with $\vartheta^{\star}=0.5\pi$ (Fig.~\eqref{fig:pullers_phase_portraits}\,(iii)), which also becomes evident in the phase portrait in Fig.~\eqref{fig:pullers_phase_portraits}\,(v). The stable configuration for trapping looking up extends across a range of initial orientations $0.3\pi\lesssim\vartheta\leq 0.5\pi$.

Strong pushers display distinct behaviors: For \,$-6\lesssim \beta \lesssim -4$, we observe oscillatory motion (Fig.~\ref{fig:phase_portrait_pusher}\,(iv)), with an amplitude that decreases with increasing  pusher strength and vanishes for $\beta\leq-7$.  This behavior appears as a limit cycle in the phase portrait, shown in Fig.~\eqref{fig:phase_portrait_pusher}\,(i) for a pusher with $\beta=-5$\,.  The oscillatory trajectory arises from the interplay between the long-ranged hydrodynamic interaction and the short-ranged repulsive force between the squirmer and the surface. More specifically, it can be understood from the perpendicular velocity $\diff h/\diff t$ and the angular velocity $\diff \vartheta/\diff t$, evaluated at the minimal and the maximum swimmer-surface distance $h=h_{\rm min}$ and $h=1.35a$, respectively (Fig.~\eqref{fig:phase_portrait_pusher} (iii)). When the pusher is at $h_{\rm min}$, $\diff h/\diff t$ has an unstable fixed point at $\vartheta\approx 0.15\pi$\,(black dashed-dotted line with the fixed-point denoted by a filled star). At this point, the orientation angle increases, $\diff\vartheta/\diff t>0$ \,(maroon dashed-dotted line). This results in a positive $\diff h/\diff t$, taking the particle away from the surface. It keeps moving away from the surface until it reaches the fixed point of the angular velocity at $h=1.35a$ and $\vartheta \approx 0.3\pi$\,(the maroon dotted line with the fixed point denoted by filled rectangle). In this configuration, the perpendicular velocity is negative and the particle moves towards the surface. This then repeats, leading to the oscillations.  

As the squirmer parameter is increased to $\beta=-10$, the limit cycle collapses to a fixed point (Fig.~\eqref{fig:phase_portrait_pusher}\,(ii)), and the pusher is sliding looking up at a steady orientation.
In this case, at $h_{\rm min}$ the fixed point of the angular velocity and the perpendicular velocity fall to the same point\,(black solid rectangle in Fig.~\eqref{fig:phase_portrait_pusher}(ii)), and the particle attains this stable orientation. It is important to note that the fixed points of $\diff h/\diff t$ and  $\diff \vartheta/\diff t$ result from the choice of the repulsive force parameters $(A,B)$ in Eq.~\eqref{eq:repulsive_zero}. A different set of parameters would dampen the oscillation for a different value of $\beta$. These findings are consistent with~\citep{Lintuvuori:2016}. 

\subsection{Corrugation-induced dynamics \label{sec:32}}
The surface corrugations change the swimmer behavior, depending on the surface wavelength $\lambda$ and the squirmer parameter $\beta$. We consider a surface of amplitude $\varepsilon=0.1$ and wavelength $\lambda=0.5-8a$, which is smaller, comparable, and larger than the size of the swimmer, respectively. In this section we focus on the dynamics arising from an initial orientation $\varphi(0)=0$ along the corrugation direction, leading to motion only in the $xz-$ plane. 

\subsubsection{Trapping of pullers at corrugation valleys}
\begin{figure}[tp]
  \centering
  \includegraphics[width=\textwidth]{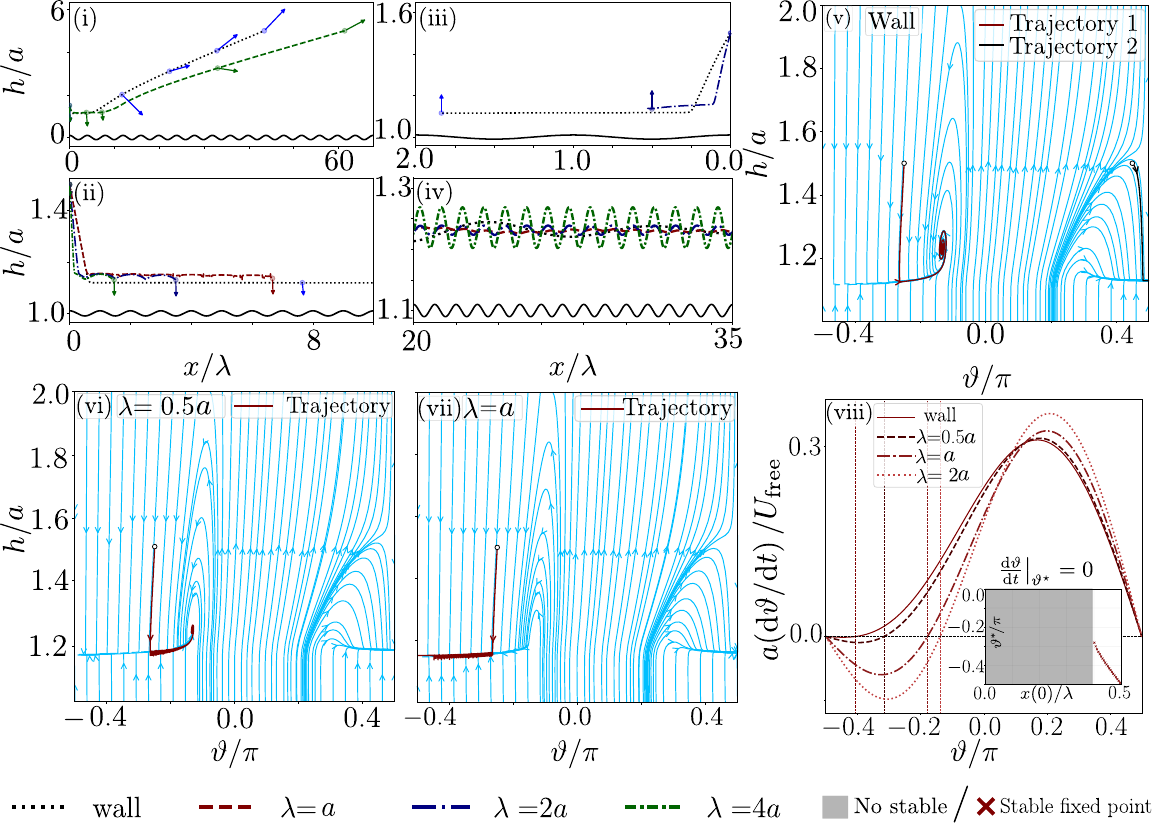}
   \caption{(i-iv) Puller trajectories near corrugated surfaces, compared to the planar wall dynamics. In panel (i) and  (ii) we consider  $\beta=3$ and $\beta=10$, respectively, with initial orientation $\left(\vartheta(0),\varphi(0)\right)=(-\pi/4,0)$ and $\lambda=2a$.  In panel (iii) we consider a puller with $\beta=10$ with initial orientation $\left(\vartheta(0),\varphi(0)\right)=(0.35\pi,0)$ near a corrugated surface with $\lambda=2a$. In panel (iv) the puller parameter is $\beta=5$ with initial orientation $\left(\vartheta(0),\varphi(0)\right)=(-0.15\pi,0)$, and we normalize the $x-$coordinates of the trajectory by the corresponding wavelength. For ease of visualization, we plot the underlying surface about $z=a$ in panels (ii) and (iii) and about $z=1.1a$ in panel (iv) and reduce it's amplitude to $10^{-2}$, in order to resolve the trajectory of the swimmer. In all the panels, we set the initial distance to $\left(x(0),y(0),h(0)\right)=(0,0,1.5a)$. Phase-portraits in $h-\vartheta$ space for a puller\,($\beta=5$) near a (v) planar wall and (vi)-(vii) wavy surfaces of wavelengths $\lambda=0.5a$ and $a$, respectively. In panels (v)-(vii) the maroon line represents the superimposed trajectory of the corresponding squirmer with $\vartheta(0)=-\pi/4$ and in panel (v) the black line represents the trajectory with $\vartheta(0)=0.45\pi$. (viii) Angular dynamics ${\rm d}\vartheta/{\rm d}t$ for different $\lambda$ at $x_S=\lambda/2$. Inset shows the fixed points $\vartheta^{\star}$ as a function of $x/\lambda$ (with $\lambda=2a$). In (v)-(viii) we consider $\varphi(0)=0.0$ and in all the panels we use $A/(\mu U_{\rm free} a)=10^7$ and $B=90 a$ for the repulsive force.}
   \label{fig:pullers_phase_portraits}
	\end{figure}

Our phase plots [Figs.~\eqref{fig:phase_plot}] show that pullers, which are initially pointing normal towards the surface, are trapped looking down, similar to their counterparts near a flat wall. 
We further observe that the scattering of weak pullers\,($0<\beta<3$) is delayed compared to their behavior near rigid walls, see Fig.~\eqref{fig:pullers_phase_portraits}\,(i).  Increasing the squirmer parameter\,($4\leq\beta\leq5$), pullers with a shallow pitch angle move parallel to the surface and exhibit oscillatory dynamics (Fig.~\eqref{fig:pullers_phase_portraits}\,(iv)). The oscillation amplitude increases with increasing wavelength~$\lambda$. For smaller wavelengths $\lambda$, the oscillations fade out and dynamics appear as sliding motion. 

In contrast to planar surfaces, for larger pitch angles strong pullers are trapped by the surface corrugations with a stable orientation of $\vartheta^{\star}=-0.5\pi$ (Figs.~\eqref{fig:phase_plot}).  Importantly, we observe that this trapping happens at the `valley' of the wavy surface, i.e, at $x(t)\, {\rm mod}\, (\lambda)=0.5\lambda$ (Fig.~\eqref{fig:pullers_phase_portraits}\,(ii)). 
Here, $h^\star=h_{\rm min}$, $\vartheta^\star=-0.5\pi$ is a stable fixed point of the angular dynamics $\mathrm{d}\vartheta/\mathrm{d}t$ (Fig.~\eqref{fig:pullers_phase_portraits}\,(viii)). The trapping region in the $\vartheta(0)-\beta$ space increases with increasing wavelength~$\lambda$. The latter can be understood by inspecting  $\mathrm{d}\vartheta/\mathrm{d}t$, whose unstable fixed point moves towards $\vartheta=0$ for increasing $\lambda$, thus increasing the range of initial pitch angles leading to trapping. The phase-portraits in  Fig.~\eqref{fig:pullers_phase_portraits}\,(vi)-(vii) illustrate this point: while for $\lambda=0.5a$ the trajectory reaches a sliding state with negligible oscillations, at $\lambda=a$ the same initial configuration leads to trapping. Lastly, we note that the stable fixed point also depends on the location with respect to the surface corrugation (inset of  Fig.~\eqref{fig:pullers_phase_portraits}\,(viii)). TOur analysis shows that there is no stable fixed point in the region of $0\leq x\lesssim 0.4\lambda$ for $\lambda=a$. Beyond this region, there is a stable orientation $\vartheta^\star$ pointing towards the surface, leading to motion along the corrugations until the squirmer reaches the `valley'\,($x=0.5\lambda$) where it becomes trapped. 

Finally, strong pullers initially pointing away from the surface ($0.2\pi\lesssim \vartheta\lesssim 0.5\pi$), are trapped looking up due to the strong hydrodynamic attraction, see Fig.~\eqref{fig:pullers_phase_portraits}\,(iii).  

	\begin{figure}[tp]
		\centering
		\includegraphics[width=\textwidth]{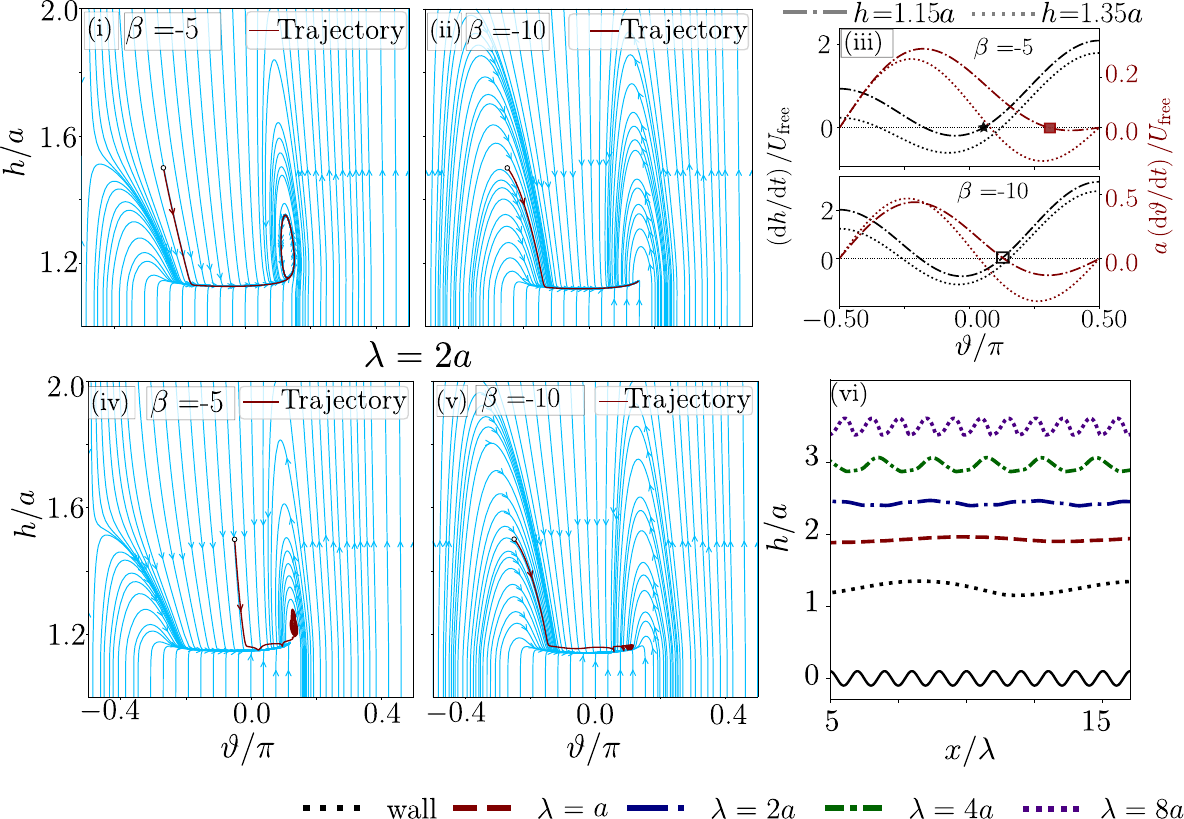}
		\caption{Phase-portraits in $h-\vartheta$ space for a pusher near a (i-ii) planar wall and (iv)-(v) wavy surfaces of wavelengths $\lambda=2a$ for $\beta=-5,10$, respectively. (iii) Angular and height dynamics, ${\rm d}\vartheta/{\rm d}t$ and ${\rm d}h/{\rm d}t$, for different distances as a function of $\vartheta$. (vi) Trajectories of squirmers  with $\beta=-5$ for different wavelengths~$\lambda$ and initial orientation $\vartheta(0)=-\pi/4$. We plot the trajectory after the swimmer has reached the oscillatory state and trajectories are shifted for better visualization. The `wall' trajectory is drawn at $h=1.2a$ and thereafter all trajectories are shifted by $0.5a$. Here, we set the initial distance to $\left(x(0),y(0),h(0)\right)=(0,0,1.5a),\,\,\varphi(0)=0$  and use $A/(\mu U_{\rm free} a)=10^7$ and $B=90 a$ for the repulsive force. For (iv), (v) and (vi) we use $\varepsilon=0.1$.}
		\label{fig:phase_portrait_pusher}
	\end{figure} 
\subsubsection{Near-surface oscillations of pushers}
Pusher squirmers with $-4<\beta<0$, initially pointing towards the surface\,($\vartheta(0)<0$) are scattered by the corrugated surface with a delay while those with $\vartheta(0)=-\pi/2$ become trapped, just as their puller counterparts. Most prominently, increasing the pusher strength $|\beta|$ and the surface wavelength $\lambda$ we discover an oscillatory regime (Fig.~\eqref{fig:phase_portrait_pusher}~(vi)). For large $\lambda=8a$, the trajectory's wavelength matches that of the surface $\lambda$. For a smaller surface wavelength $\lambda=2a$, however, we identify two characteristic wavelengths in the trajectory: the wavelength corresponding to the underlying surface and that due to the oscillations arising from the repulsive force. Decreasing $\lambda$ further, leads to oscillatory dynamics at the same wavelength observed near a planar wall. Interestingly, the amplitude of the oscillations also varies with $\lambda$ and it appears to be smallest at an intermediate $\lambda$, where both wavelengths are apparent in the trajectory. 

The oscillatory dynamics, resulting from the interplay of reorientation and hydrodynamic attraction, can be quantfied in phase portraits [Fig.\,~\eqref{fig:phase_portrait_pusher} (vi)]. These highlight the change of the limit cycle radius induced by the surface corrugations, compared to the planar wall results. They further show that strong pushers ($\beta=-10$) start oscillating, albeit with a small amplitude. 

Lastly, we note that the behavior of neutral squirmers ($\beta=0$) merely changes due to surface corrugations and they scatter similarly to weak pushers and pullers. 

\subsubsection{Flow fields of squirmers near surface corrugations}
\begin{figure}[htp]
	\centering
    \includegraphics[width=\textwidth]{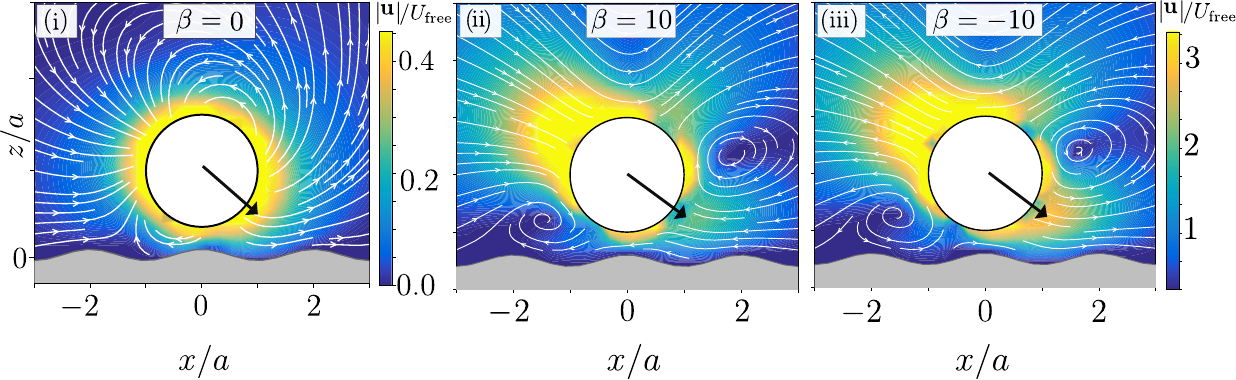}
	\caption{Flow fields of (i)~a neutral squirmer (ii)~a puller and (iii)~a pusher near a surface corrugation with $\lambda=2a$ and amplitude $\varepsilon=0.1$, obtained using MRS. The swimmer is at $h=1.5a$ and at an orientation $(\vartheta,\varphi)=(-\pi/5,0)$, indicated by the black arrow. The scale bars show the velocity magnitude $|\vec{u}|/U_{\rm free}$, rescaled by the swim speed in free space. The color bar for (ii) is the same as in (iii).}
	\label{fig:first_order_flow_field}
\end{figure}

To rationalize our main findings -- scattering of neutral squirmers, trapping of pullers, and oscillating pushers -- we  obtain the flow fields produced by squirmers near the corrugated surface using MRS. A neutral squirmer, shown in Fig.~\eqref{fig:first_order_flow_field}\,(i), pushes the fluid from the front, while pulling it from the back. While it is pointing towards the surface, the fluid pushed from the front has to turn to satisfy the no-slip boundary condition on the surface. This results in a reorientation of the squirmer away from the surface, where the fluid pushed from the front has a much larger space. The turning of the neutral squirmer away from the surface results in it being scattered. 

The puller squirmer\,($\beta=10$), shown in Fig.~\eqref{fig:first_order_flow_field}\,(ii), pulls the fluid from the front and the back and pushes the fluid from the sides. While it is pointing towards the surface, the fluid pushed from either side experiences different hydrodynamic interactions with the surface. The fluid pushed from the squirmer side closer to the surface has to turn in a small space compared to the side farther from the surface. This creates an imbalance in the reorientation velocity and the swimmer reorients to counteract the imbalance. Two symmetric configurations represent the squirmer either pointing down on top of the `hill' or the `valley'. At the `hill' the puller is in an unstable configuration as a small perturbation of its position towards either side would create more space for the pushed fluid, resulting in the puller `falling' along the surface. At the top of the `valley' the puller is stable as `climbing' along the surface would result in a reorientation due to smaller space available along the `climbing' direction. Thus the squirmer is `trapped looking down' at the `valley'. As discussed above, this trapping increases with the strength of the puller as the flow pushed from the side has a larger velocity and hence the fluid has to turn faster near the surface, leading to a faster reorientation velocity. The `trapping' is more significant for larger wavelength due to the larger space available in the `valley'. For wavelengths comparable to the size of the puller, the space available for the flow is effectively reduced and the swimmer experiences an `effective' flat surface. This leads to a sliding motion of pullers\,($4\leq\beta\leq6$), similar to their observed behavior near a flat wall.

The reverse behavior is observed for the pusher \,($\beta=-10$), see Fig.~\eqref{fig:first_order_flow_field}\,(iii). The swimmer pushes the fluid from the front and the back and pulls it from the side. In an orientation where it is pointing towards the surface, the flow from the front of the swimmer has to turn in a small space, compared to when it is parallel to the surface. Thus, the pusher prefers the configuration where it is parallel or slightly looking away from the surface. In this configuration, for wavelengths of the underlying surface larger than the swimmer size, the swimming velocities are changed depending on the relative position with respect to the surface, thus leading to an oscillatory trajectory that has the signature of the underlying wavelength of the surface. For wavelengths smaller than the swimmer size, this effect is negligible as the squirmer experiences an effective averaged flow from the underlying surface, and hence the behavior is similar to that of a pusher near a planar wall, with sliding motion looking away from the surface.

\subsection{Three--dimensional motion}
\begin{figure}[htp]
	\centering
    \includegraphics[width=\textwidth]{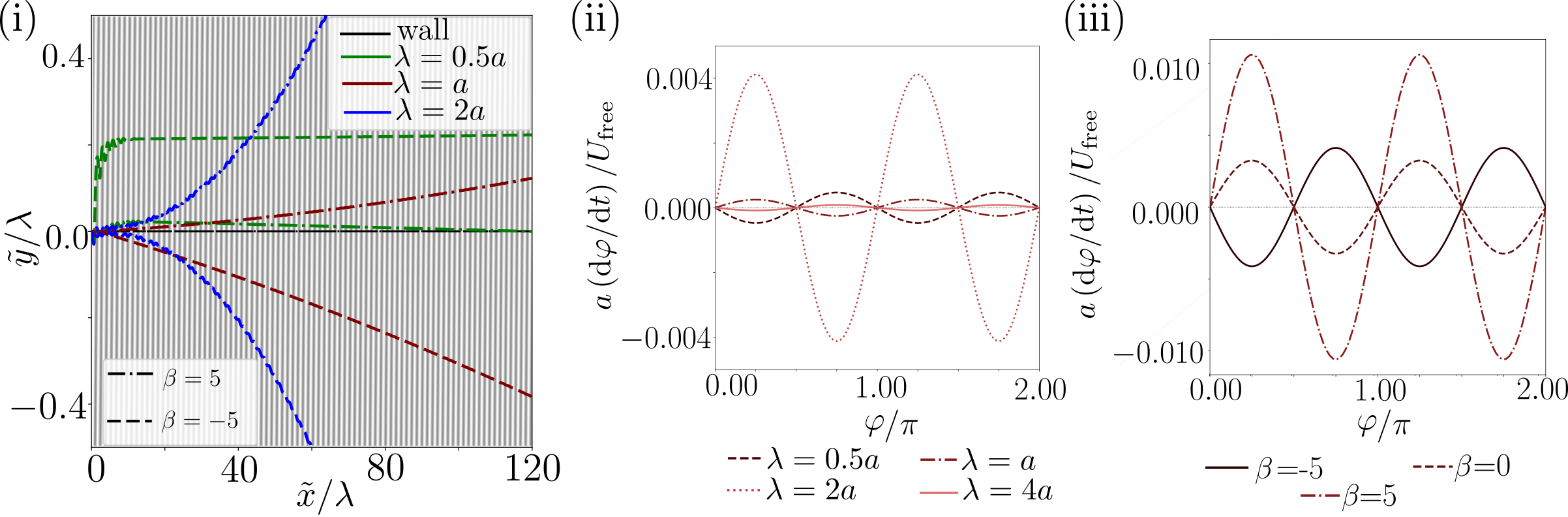}
	\caption{(i)2D-projection of squirmer dynamics near a corrugated surface. The background indicates the surface heights (dark) and valleys (white). Squirmers are initially oriented at an angle $\varphi(0)=\pi/4$ with respect to the surface corrugations. Different colors indicate different wavelength, while dashed and dashed-dotted lines correspond to pushers and pullers, respectively. The trajectories are plotted in $(\tilde{x},\tilde{y})\equiv\left(x\cos\varphi(0)+y\sin\varphi(0),-x\sin\varphi(0)+y\cos\varphi(0)\right)-$space and rescaled with respect to the respective surface wavelength. Angular dynamics $\diff\varphi/\diff t$ with respect to $\varphi$ for (ii) different wavelengths (and $\beta=5$) and (iii) different squirmer parameters\,(and $\lambda=2a$) . The other parameters used in (i) are $(x(0),y(0),h(0))=(0,0,1.5a)$, $\vartheta(0)=-0.15\pi$, $\varepsilon=0.1$ and we use $A/(\mu U_{\rm free} a)=10^7$ and $B=90 a$ for the repulsive force. We evaluate (ii) and (iii) at $(x,y,h)=(0,0,1.22a)$,  with $\vartheta=-0.13\pi$ and $\varepsilon=0.1$.}
	\label{fig:3d_trajectories}
\end{figure}

We first inspect squirmers with an initial orientation perpendicular to the corrugation direction\,($\varphi=0.5\pi$) (or equivalently, aligned with the hills and valleys). Squirmers starting on top of a hill, $x/\lambda=0$, have a behavior similar to that of the corresponding squirmer near a planar wall, since the swimmer experiences no asymmetry in the underlying surface. Neutral squirmers behave similar to pullers, but  are scattered by the surface, irrespective of~$\varphi(0)$. Interesting behaviors emerge, as we tilt the initial orientation $\varphi(0)=\pi/4$ with respect to the corrugations, so that it is neither aligned with the corrugation direction ($x-$axis; $\varphi(0)=0$) nor perpendicular to it ($y-$axis; $\varphi(0)=\pi/2$). We consider parameter regimes, where we identified near-surface (oscillatory or sliding) motion for both pushers and pullers with~$\varphi(0)=0$ (see Fig.~\eqref{fig:phase_plot}). Most prominently, in addition to changing their pitch angle $\vartheta$ pushers and pullers change their in-plane orientation  $\varphi$, leading to an in-plane motion.  This depends both on the wavelength $\lambda$ and the squirmer parameter $\beta$. While it is observed that the puller ($\beta=5$) near a surface of wavelength $\lambda=0.5a$ turns towards the negative $\tilde{y}$ direction, for wavelengths $\lambda=a,2a$ it moves towards the positive $\tilde{y}$ direction. To understand this behavior, we study the angular velocity $\diff\varphi/\diff t$ for surface wavelengths $\lambda=0.5a,a,2a$ and $=4a$ (Fig.~\eqref{fig:3d_trajectories}\,(ii)). Its magnitude is small compared to the perpendicular velocity $\diff h/\diff t$ and the angular velocity $\diff \vartheta/\diff t$ and thus evaluation of $\diff \varphi/\diff t$ is sufficient to understand the drift direction. (We evaluate it at the average height $h$ and orientation $\vartheta$ of the squirmer.) For all wavelengths $\lambda$, we find that the fixed point is $\phi=n\pi/2$, where $n$ is an integer, which results from the fact that the swimmer experiences a symmetric surface when oriented along $\varphi=n\pi/2$. The stability of the fixed points, however, depends on the wavelength: For $\lambda=0.5a, 4a$, $\varphi=0.5\pi$ is an unstable fixed point, which becomes stable for $\lambda=a$ and $=2a$. Thus, the trajectories of pullers initially oriented along $\varphi=\pi/4$ will ultimately follow the $x-$ or $y-$ direction, depending non-trivially on the wavelength $\lambda$. For pushers ($\beta=-5$) this behavior reverses, due to a change of stability of the fixed points~(Fig.~\eqref{fig:3d_trajectories}\,(iii)). 

Lastly, we note that the qualitative behavior of trapping of strong pullers as well as the scattering of weak squirmers, including neutral ones, persist also for varying $\varphi(0)$.  

\section{Discussion and conclusions\label{sec:4}} Employing a perturbative scheme based on a bispherical approach, we investigate the dynamics of squirmers near a periodic surface. Most prominently, our findings reveal two phases emerging from the interplay of short-range repulsive and hydrodynamic couplings with the surface corrugations that are distinct from squirmer behaviors near planar walls: Pullers become trapped at surface valleys, while pushers exhibit oscillatory dynamics in phase with the underlying surface structure. Moreover, the scattering of squirmers from the surface becomes delayed by the corrugations. We further demonstrate that -- depending on the relative angle between squirmers and corrugations -- surface structure can sort pushers from pullers by inducing a wavelength-dependent reorientation. 

Our newly-identified trapping phase of pullers qualitatively matches recent experiments of {\it Chlamydomonas reinhardtii}, which were observed to spend significantly more time in the valleys of surface corrugations than near the hills~\citep{Li:2026}. While the near-field flows of the algae differ from the paradigmatic squirmer model, our findings suggest that puller-type flow fields contribute to trapping in these microstructures. We further note that oscillatory dynamics of squirmers near periodic surfaces have been found in recent theoretical work on squirmers~\citep{Ishimoto:2023}. Our work extends these findings by explicitly introducing steric repulsion, which together with our perturbative scheme provides access to the full phase diagram. It further lays the foundation for studying more complex surface structures, allowing to identify topographies for optimal microswimmer guidance. Our semi-analytical framework can be readily applied to investigate the role of near-field effects on surface deformations, relevant in the context of microswimmer-membrane interactions~\citep{Vutukuri:2020,Takatori:2020}, and the resultant dynamical behaviors~\citep{Garai:2025}. Relaxing the constraint of small-surface amplitude, represents an interesting future research endeavor, relying on the development of numerical schemes, such as the MRS used to validate our perturbative calculations. 

It is important to add that wall interactions drastically change in the presence of external flows, where fluid vorticity reorients active agents, leading to upward swimming (i.e. rheotaxis)~\citep{Mathijssen:2019} and accumulation behind surface corrugations~\citep{Secchi:2020, Pellegrino:2026}. Our approach can be extended to include shear flows~\citep{Ghosh:2023, Kurzthaler:2024:JFM}, allowing to probe the interplay of flow vorticity and near-field hydrodynamic interactions with the surface corrugations. 

While our work focuses on spherical microswimmers, the aspect of shape represents an important ingredient for surface-interactions and swimmer-elongation is expected to lead to new dynamical behaviors. Indeed, recent work has demonstrated that the hydrodynamic coupling between the time-dependent but reversible swim stroke of elongated ciliated swimmers and the surface corrugations of the enclosing channel can induce directed motion~\citep{Antunes:2026}. Understanding the approach of these elongated swimmers towards the corrugated surface, however, remains an open research question. Interestingly, explicit simulations of flagellated bacteria, using multi-particle collision dynamics, demonstrate oscillatory  dynamics near corrugated surfaces, in accord with our observations for pushers~\citep{Martin:2026}. At large surface curvatures, the torque-dipole resulting from the rotation of the flagella bundle also leads to circular in-plane motion with a sense of rotation opposite to the planar wall. These findings highlight the importance of three-dimensional surface structure on microbial dynamics and suggest that the latter can crucially modify microbial phenomena, such as surface accumulation~\citep{Yeo:2026} and the formation of bacterial communities~\cite{Secchi:2020}.  
  
\begin{acknowledgments}
    We are grateful for discussions with Matteo Ciarchi, Mirko Residori, Yanis Baouche, and Akhil Varma. We thank Dana Ferranti for helpful discussions and sharing code for the method of regularized Stokeslet surfaces. TGF acknowledges support from National Science Foundation (NSF) grant DMS-2512565.
\end{acknowledgments}

\section*{Conflict disclosure} 
There are no conflicts to declare.

%
%
%
%
%
%

\appendix
\section{Details on the calculation of the swimming velocity in bispherical coordinates \label{appA}}
Here, we focus on computing the velocity fields ($\unitvec{u}$, $\vec{u}^{(0)}$) and pressure fields ($\hat{p}$, $p^{(0)}$) using bispherical coordinates by assuming that the boundary conditions on the surface of the sphere (or swimmer) are known. In practice, for the auxiliary problem we impose unit $\unitvec{U}$ and $\unitvec{\Omega}$ to compute the generalized resistance matrices $\unitvec{\mathcal{T}}$ and $\unitvec{\mathcal{R}}$ as input for Eq.~\eqref{eq:vel_correction_repulsive}. For the swimmer case, we first compute $\vec{U}^{(0)}$ and $\vec{\Omega}^{(0)}$ using  Eq.~\eqref{eq:vel_correction_repulsive} for $\varepsilon=0$ and then compute $\vec{u}_{\mathcal{S}_M}$ by evaluating the $\partial_z\vec{u}^{(0)}|_{z=0}$ as input for $\vec{u}_{\mathcal{S}_M}$. 

In the following presentation we denote the velocities by $\vec{u}$ and pressure by $p$. For the sake of completeness we outline here the projection of the incompressibility condition and the zeroth-order boundary conditions on the bispherical basis which generates a set of linear relations between the eight coefficients\,$\{A_{n}^{m},B_{n}^{m},C_{n}^{m},D_{n}^{m},E_{n}^{m},F_{n}^{m},G_{n}^{m},H_{n}^{m}\}$~\citep{Lee:1980,Mozaffari:2016}. Here,  the value of $m$ is determined by the boundary conditions and $n\in[m,N]$, where we choose $N$ such that the value of the coefficients are negligible for $n>N$. 

\subsection{Incompressibility condition}
The incompressibility condition\,($\boldsymbol{\nabla}\cdot\boldsymbol{u}=0$) in cylindrical coordinates is expressed as
\begin{subequations}
 \begin{align}
    \frac{1}{\rho}\frac{\partial(\rho u)}{\partial \rho}+\frac{1}{\rho}\frac{\partial v}{\partial \phi}+\frac{\partial w}{\partial z}=0
\end{align} 
where $(\rho,\phi,z)$ are the cylindrical coordinates with the origin at $(x_S, y_S,0)$ in Fig.~\eqref{fig:schematic}\,(i). Substituting for the velocities from Eqs.~\eqref{eq:pressure}-\eqref{eq:w}, we note that the different $m-$modes in the azimuthal coordinate $\phi$ are decoupled. Simplifying, we obtain
\begin{align}\label{eq:continuity_0}
\left(3+\rho\frac{\partial }{\partial \rho}+z\frac{\partial }{\partial z}\right)p_{0}+2\left(\frac{1}{\rho}+\frac{\partial }{\partial \rho}\right)u_{0}+2\frac{\partial w_{0}}{\partial z}=0\quad{\rm for}\quad m=0,
\end{align}
and
\begin{align}\label{eq:continuity_m}
\left(3+\rho\frac{\partial }{\partial \rho}+z\frac{\partial }{\partial z}\right)p_m+\left(\frac{\partial }{\partial \rho}+\frac{m+1 }{\rho}\right)\gamma_m+\left(\frac{\partial }{\partial \rho}-\frac{m-1 }{\rho}\right)\chi_m+2\frac{\partial w_m}{\partial z}=0\quad{\rm for}\quad m\geq1.
\end{align}
respectively.
\end{subequations}
Substituting Eqs.~\eqref{eq:u0}-~\eqref{eq:w_m} in Eq.~\eqref{eq:continuity_0}\,(or Eq.~\eqref{eq:continuity_m}), we obtain for $m=0$
\begin{subequations}
\begin{align}
-\frac{n}{2}A_{n-1}^{0}&+\frac{5}{2}A_{n}^{0}+\frac{1}{2}(n+1)A_{n+1}^{0}-nD_{n-1}^{0}+(2n+1)D_{n}^{0}-(n+1)D_{n+1}^{0}\nonumber\\&-n(n-1)E_{n-1}^{0}+2n(n+1)E_{n}^{0}-(n+1)(n+2)E_{n+1}^{0}=0,\label{in1m0}
\end{align}
and
\begin{align}
-\frac{n}{2}B_{n-1}^{0}&+\frac{5}{2}B_{n}^{0}+\frac{1}{2}(n+1)B_{n+1}^{0}-nC_{n-1}^{0}+(2n+1)C_{n}^{0}-(n+1)C_{n+1}^{0}\nonumber\\&-n(n-1)F_{n-1}^{0}+2n(n+1)F_{n}^{0}-(n+1)(n+2)F_{n+1}^{0}=0,\,\,\label{in2m0}
\end{align}
\end{subequations}
by projecting on $(\cosh\eta-\mu)^{1/2}\cosh\left\{(n+1/2)\eta\right\}P_n(\mu)$ and $(\cosh\eta-\mu)^{1/2}\sinh\left\{(n+1/2)\eta\right\}P_n(\mu)$ respectively.
Similarly, for $m\geq1$ by projecting on $(\cosh\eta-\mu)^{1/2}\cosh\left\{(n+1/2)\eta\right\}P_n^{m}(\mu)$ and $(\cosh\eta-\mu)^{1/2}\sinh\left\{(n+1/2)\eta\right\}P_n^{m}(\mu)$, we obtain
\begin{subequations}
\begin{align}
-\frac{1}{2}(n-m)A_{n-1}^{m}+\frac{5}{2}A_{n}^{m}+\frac{1}{2}(n+m+1)A_{n+1}^{m}-(n-m)D_{n-1}^{m}&+(2n+1)D_{n}^{m}-(n+m+1)D_{n+1}^{m}\nonumber\\-\frac{1}{2}(n-m)(n-m-1)E_{n-1}^{m}+(n-m)(n+m+1)E_{n}^{m}&-\frac{1}{2}(n+m+1)(n+m+2)E_{n+1}^{m}\nonumber\\&+\frac{1}{2}G_{n-1}^{m}-G_{n}^{m}+\frac{1}{2}G_{n+1}^{m}=0\label{in1m}
\end{align}
and
\begin{align}
-\frac{1}{2}(n-m)B_{n-1}^{m}+\frac{5}{2}B_{n}^{m}+\frac{1}{2}(n+m+1)B_{n+1}^{m}-(n-m)C_{n-1}^{m}&+(2n+1)C_{n}^{m}-(n+m+1)C_{n+1}^{m}\nonumber\\-\frac{1}{2}(n-m)(n-m-1)F_{n-1}^{m}+(n-m)(n+m+1)F_{n}^{m}&-\frac{1}{2}(n+m+1)(n+m+2)F_{n+1}^{m}\nonumber\\&+\frac{1}{2}H_{n-1}^{m}-H_{n}^{m}+\frac{1}{2}H_{n+1}^{m}=0,\label{in2m}
\end{align}
\end{subequations}
respectively.
\subsection{Wall boundary condition}
The no-slip boundary condition on the wall $\mathcal{S}_0$\,($z=0$ or $\eta=0$) is projected on the bispherical basis by substituting Eqs.~\eqref{eq:u}-~\eqref{eq:w} and Eqs.~\eqref{eq:u0}-~\eqref{eq:w_m} in Eq.~\eqref{eq:zeroth_bc_1}, which generates
\begin{subequations}
\begin{align}
w=0,\,\,&{\rm or}\,\,D_{n}^{m}=0\,\,\forall n,m,\label{bibc11}\\
u=0,\,\,v=0\,\,{\rm or}\,\,u_{0}=-\frac{\rho}{2}p_{0}, \,v_{0}=0,{\rm for}\,\,m=0,\,\,&{\rm and}\,\,\gamma_{m}=\chi_{m}=-\frac{\rho}{2}p_{m},\,\,{\rm for}\,\,m\geq1.\label{bibc12}
\end{align}
Eqs.~\eqref{bibc11}-~\eqref{bibc12} are satisfied for all $\eta$ and $\mu$ if
\begin{align}
-\frac{(n-m-1)}{2n-1}F_{n-1}^{m}+F_{n}^{m}&-\frac{(n+m+2)}{2n+3}F_{n+1}^{m}-\frac{1}{2(2n-1)}B_{n-1}^{m}+\frac{1}{2(2n+3)}B_{n+1}^{m}=0,\,\,\forall m,\label{bibc1con}\\
&H_{n}^{0}=0,\,\,{\rm for}\,\,m=0,\,\,{\rm and }\label{bibc2con}\\
-\frac{(n-m+1)}{2n-1}H_{n-1}^{m}+H_{n}^{m}&-\frac{(n+m)}{2n+3}H_{n+1}^{m}+\frac{(n-m)(n-m+1)}{2(2n-1)}B_{n-1}^{m}-\frac{(n+m)(n+m+1)}{2(2n+3)}B_{n+1}^{m}\nonumber\\&\quad=0\quad{\rm for}\,\, m\geq1.\label{bibc3con}
\end{align}   
\end{subequations}
\subsection{Sphere boundary condition}
The boundary condition on the sphere surface $\mathcal{S}_{P}$ is satisfied by expressing the right hand side of Eq.~\eqref{eq:zeroth_bc_2} in terms of bispherical coordinates. We express the velocity on the surface of the sphere as $\vec{u}=u_s\unitvec{\rho}+v_s\unitvec{\phi}+w_s\unitvec{z}$, with 
expansion in bispherical coordinates:
\begin{subequations}
\begin{align}
u_{s}&=\sum_{m}u_{s}^{m}(\eta,\mu)\cos(m\phi+\alpha_{m}),\label{bibc21}\\
v_{s}&=\sum_{m}v_{s}^{m}(\eta,\mu)\sin(m\phi+\alpha_{m}),\label{bibc22}\\
w_{s}&=(\cosh \eta_{0}-\mu)^{\frac{1}{2}} \sum_{m=0}\sum_{n=m}Z_{n}^{m}P_{n}^{m}(\mu)\cos(m\phi+\alpha_{m}).\label{bibc23}
\end{align}   
For $m=0$, Eqs.~\eqref{bibc22}-~\eqref{bibc23} is expanded as
\begin{align}
u_{s}^{0}=(\cosh \eta_{0}-\mu)^{\frac{1}{2}}\sum_{n=1}X_{n}^{0}P_{n}^{1}(\mu),\label{bibc24}\\
v_{s}^{0}=(\cosh \eta_{0}-\mu)^{\frac{1}{2}}\sum_{n=1}Y_{n}^{0}P_{n}^{1}(\mu),\label{bibc25}
\end{align}
while for $m\geq1$ we write
\begin{align}
u_{s}^{m}+v_{s}^{m}=(\cosh \eta_{0}-\mu)^{\frac{1}{2}}\sum_{n=m+1}X_{n}^{m}P_{n}^{m+1}(\mu),\label{bibc26}\\
u_{s}^{m}-v_{s}^{m}=(\cosh \eta_{0}-\mu)^{\frac{1}{2}}\sum_{n=m-1}Y_{n}^{m}P_{n}^{m-1}(\mu).\label{bibc27}
\end{align}
\end{subequations}
Here $\{X_n^m,Y_n^m,Z_n^m\}$ are known through the expansion of the prescribed slip velocity and the rigid body velocity on the surface of the sphere. Substituting Eqs.~\eqref{eq:u0}-~\eqref{eq:w_m} into Eqs.~\eqref{bibc24}-~\eqref{bibc27}, and then  comparing the coefficients of the Legendre polynomials $P_n^m(\mu)$ leads to
\begin{subequations}
\begin{align}
E_{n}^{m} \sinh \left(n+\frac{1}{2}\right)\eta_{0}+F_{n}^{m} \cosh \left(n+\frac{1}{2}\right)\eta_{0}-\frac{1}{\sinh \eta_{0}}\left[\frac{1}{(2n+3)}C_{n+1}^{m}\sinh \left(n+\frac{3}{2}\right)\eta_{0}\right.\nonumber\\
\left.-\frac{1}{(2n-1)}C_{n-1}^{m}\sinh \left(n-\frac{1}{2}\right)\eta_{0}\right]=X_{n}^{m}(\eta_{0})-\frac{1}{\sinh\eta_0}\left[\frac{1}{(2n+3)}Z_{n+1}^{m}(\eta_{0})-\frac{1}{(2n-1)}Z_{n-1}^{m}(\eta_{0})\right],\,\,\forall m \label{bibc2coeff1}\\
G_{n}^{0}\sinh \left(n+\frac{1}{2}\right)\eta_{0}+H_{n}^{0}\cosh \left(n+\frac{1}{2}\right)\eta_{0}=Y_{n}^{0}(\eta_{0}),\,\,{\rm for}\,\,m=0 \label{bibc2coeff2}\\
G_{n}^{m} \sinh \left(n+\frac{1}{2}\right)\eta_{0}+H_{n}^{m} \cosh \left(n+\frac{1}{2}\right)\eta_{0}+\frac{1}{\sinh \eta_{0}}\left[\frac{(n+m)(n+m+1)}{(2n+3)}C_{n+1}^{m}\sinh \left(n+\frac{3}{2}\right)\eta_{0}\right.\nonumber
\\\left.-\frac{(n-m)(n-m+1)}{(2n-1)}C_{n-1}^{m}\sinh \left(n-\frac{1}{2}\right)\eta_{0}\right]=Y_{n}^{m}(\eta_{0})+\frac{1}{\sinh\eta_0}\left[\frac{(n+m)(n+m+1)}{(2n+3)}Z_{n+1}^{m}(\eta_{0})\right.\nonumber
\\\left.-\frac{(n-m)(n-m+1)}{(2n-1)}Z_{n-1}^{m}(\eta_{0})\right],\,\,{\rm for}\,\, m\geq1 \label{bibc2coeff3}
\end{align}  
\begin{align}
A_{n}^{m} \sinh \left(n+\frac{1}{2}\right)\eta_{0}+B_{n}^{m} \cosh \left(n+\frac{1}{2}\right)\eta_{0}-\frac{2}{\sinh \eta_{0}}\left[\frac{(n+m+1)}{(2n+3)}C_{n+1}^{m}\sinh \left(n+\frac{3}{2}\right)\eta_{0}\right.\nonumber\\\left.-\cosh \eta_{0}C_{n}^{m}\sinh\left(n+\frac{1}{2}\right)\eta_{0}+\frac{(n-m)}{(2n-1)}C_{n-1}^{m}\sinh \left(n-\frac{1}{2}\right)\eta_{0}\right]=\nonumber\\
-\frac{2}{\sinh\eta_{0}}\left[\frac{(n+m+1)}{(2n+3)}Z_{n+1}^{m}(\eta_{0})-\cosh\eta_{0}Z_{n}^{m}(\eta_{0})+\frac{(n-m)}{(2n-1)}Z_{n-1}^{m}(\eta_{0})\right]\,\,\forall m. \label{bibc2coeff4}
\end{align}
\end{subequations}

The series expansions in eigenfunctions of the bispherical basis are infinite and truncated for some value of $n=N$. The error in truncating the series for some finite $N$ is estimated in Ref.~\citep{Lee:1980}.

From Eq.~\eqref{bibc11}, it is evident that there are seven independent coefficients\,$\{A_{n}^{m},B_{n}^{m},C_{n}^{m},E_{n}^{m},F_{n}^{m},G_{n}^{m},H_{n}^{m}\}$, where for $A_{n}^{m},B_{n}^{m}$ and $C_{n}^{m}$, $n$ ranges from $m$ to $N$, thus having $3(N-m+1)$ coefficients. For the coefficients $E_{n}^{m}$ and $F_{n}^{m}$, $n$ ranges from  $m+1$ to $N$ thus having $(N-m)$ independent coefficients. For the coefficients $G_{n}^{0}$ and $H_{n}^{0}$ $n$ ranges from $1$ to $N$ and for coefficients $G_{n}^{m}$ and $H_{n}^{m}$\,($m\geq1$) $n$ ranges from $m-1$ to $N$ , thus having a total of $2N$ and $2(N-m+2)$ coefficients respectively for $m=0$ and $m\geq1$ respectively. Thus, we have a total of $7N+3$ and $7(N-m+1)$ for $m=0$ and $m\geq1$, respectively. For $m=0$, the number of independent equations are $2(N+1)$ from Eqs.~\eqref{in1m0} and~\eqref{in2m0}, $2N$ from Eqs.~\eqref{bibc1con} and~\eqref{bibc2con} , and $2N$ and $(N+1)$ equations from Eqs.~\eqref{bibc2coeff1},~\eqref{bibc2coeff2} and~\eqref{bibc2coeff4}, respectively, thus having a total of $7N+3$ equations. Similarly for $m\geq1$, the number of independent relations are $2(N-m+1)$ from Eqs.~\eqref{in1m} and~\eqref{in2m}, $(N-m)$  and $(N-m+2)$ from Eqs.~\eqref{bibc1con}) and~\eqref{bibc3con}, respectively , and $N-m$, $N-m+2$ and  $N-m+1$ equations from Eqs.~\eqref{bibc2coeff1},~\eqref{bibc2coeff3} and~\eqref{bibc2coeff4} respectively, thus having a total of $7(N-m+1)$ equations.  The constants $\{X_{n}^{m},Y_{n}^{m},Z_{n}^{m}\}$ are obtained by expanding the sphere boundary condition in bispherical eigenfunctions as in Eqs.~\eqref{bibc21}-\eqref{bibc23}. Thus we obtain a system of linear equations of the form 
\begin{align}
\mathbf{A}\boldsymbol{x}=\boldsymbol{b},
\end{align}
where $\mathbf{A}$ is the $7(N-m+1)\times 7(N-m+1)$ for $m\geq1$ \,(or  $7N+3\times 7N+3$ for $m=0$) banded coefficient matrix obtained from the boundary conditions, $\boldsymbol{x}$ is the $7(N-m+1)$ for $m\geq1$ \,(or  $7N+3$ for $m=0$) vector of the seven independent coefficients\,$\{A_{n}^{m},B_{n}^{m},C_{n}^{m},E_{n}^{m},F_{n}^{m},G_{n}^{m},H_{n}^{m}\}$, and $\boldsymbol{b}$ is the vector $\{X_{n}^{m},Y_{n}^{m},Z_{n}^{m}\}$ obtained from the sphere boundary conditions of the same dimensions as that of $\boldsymbol{x}$. The number of coefficients and equations are balanced and hence can be solved using a matrix solver numerically. For our calculation we LU decompose the individual $m-$ blocks of the $\mathbf{A}$ matrix using the ``lu$\_$factor" function from ``scipy$.$linalg" and then use the ``lu$\_$solve" to solve for the coefficients.

\subsection{Projection of the squirming boundary conditions in bispherical basis}\label{squirmerbc}
The squirmer slip velocity at the surface of the sphere\,$\mathcal{S}_{P}$\,($\eta=\eta_0$) is analytically expanded in bispherical coordinates using the relations in\,Eqs.~\eqref{bibc21}-~\eqref{bibc27}. Equation~\eqref{eq:squirmer} is expressed in cylindrical coordinates with the origin at $(x_S, y_S, 0)$\,(see Fig.~\eqref{fig:schematic}). Through the transformation of the unit vectors using $\boldsymbol{\hat{x}}=\cos \phi \boldsymbol{\hat{\rho}}-\sin \phi\boldsymbol{\hat{\phi}}$, $\boldsymbol{\hat{y}}=\sin \phi \boldsymbol{\hat{\rho}}+\cos \phi\boldsymbol{\hat{\phi}}$ and $\boldsymbol{\hat{z}}=\boldsymbol{\hat{z}}$, the orientation vector is expressed as
\begin{align}
	\boldsymbol{e}(t)=\cos\vartheta(t)\left\{\cos(\varphi(t)-\phi) \boldsymbol{\hat{\rho}}+ \sin (\varphi(t)-\phi)\right\} \boldsymbol{\hat{\phi}}+\sin\vartheta(t)\boldsymbol{\hat{z}}.
\end{align}
In cylindrical coordinates, rewriting $\boldsymbol{r}'=\left\{\rho \boldsymbol{\hat{\rho}}+\left(z-h(t)\right)\boldsymbol{\hat{z}}\right\}/a$, which is the vector pointing from the center of the squirmer to its surface, Eq.~\eqref{eq:squirmer} is expressed as
\begin{align}
\boldsymbol{u}_{s}=B_{1}(\boldsymbol{u}_{B_{1}}
+\beta \boldsymbol{u}_{B_{2}}). \label{eq:squirmer_modes_expansion}
\end{align}
For projecting the squirmer slip velocity in bispherical coordinates we express Eq.~\eqref{eq:squirmer_modes_expansion} in terms of its azimuthal modes as
\begin{subequations}
\begin{align}
\boldsymbol{u}_{B_{1}}\equiv &\frac{1}{a^2}\left[\left\{u^{(0)}_{B_1}\sin\vartheta(t)+u^{(1)}_{B_1}\cos\vartheta(t)\cos\left(\phi-\varphi(t)\right)\right\}\unitvec{\rho}+\left\{v^{(0)}_{B_1}\sin\vartheta(t)+v^{(1)}_{B_1}\cos\vartheta(t)\sin\left(\phi-\varphi(t)\right)\right\}\unitvec{\phi}\right.\nonumber\\&\left.+\left\{w^{(0)}_{B_1}\sin\vartheta(t)+w^{(1)}_{B_1}\cos\vartheta(t)\cos\left(\phi-\varphi(t)\right)\right\}\unitvec{z}\right],\quad{\rm and} \label{eq:squimerB1} \\
\boldsymbol{u}_{B_{2}}\equiv &\frac{1}{a^3}\left[\left\{u^{(0)}_{B_2}\left(\sin^2\vartheta(t)-\frac{\cos^2\vartheta(t)}{2}\right)+u^{(1)}_{B_2}\frac{\sin 2\vartheta(t)}{2}\cos\left(\phi-\varphi(t)\right)+u^{(2)}_{B_2}\frac{\cos^2 \vartheta(t)}{2}\cos 2\left(\phi-\varphi(t)\right)\right\}\unitvec{\rho}\right.\nonumber\\
&\left.+\left\{v^{(0)}_{B_2}\left(\sin^2\vartheta(t)-\frac{\cos^2\vartheta(t)}{2}\right)+v^{(1)}_{B_2}\frac{\sin 2\vartheta(t)}{2}\sin\left(\phi-\varphi(t)\right)+v^{(2)}_{B_2}\frac{\cos^2 \vartheta(t)}{2}\sin 2\left(\phi-\varphi(t)\right)\right\}\unitvec{\phi}\right.\nonumber\\
&\left.+\left\{w^{(0)}_{B_2}\left(\sin^2\vartheta(t)-\frac{\cos^2\vartheta(t)}{2}\right)+w^{(1)}_{B_2}\frac{\sin 2\vartheta(t)}{2}\cos\left(\phi-\varphi(t)\right)+w^{(2)}_{B_2}\frac{\cos^2 \vartheta(t)}{2}\cos2\left(\phi-\varphi(t)\right)\right\}\unitvec{z}\right], \label{eq:squimerB2}
\end{align}
where
\begin{align}
u^{(0)}_{B_1}=\rho\left\{z-h(t)\right\};\quad v^{(0)}_{B_1}&=0;\quad w^{(0)}_{B_1}= -\rho^{2},\\
u^{(1)}_{B_1}= -\left\{z-h(t)\right\}^{2};\quad v^{(1)}_{B_1}&=a^2;\quad w^{(1)}_{B_1}=\rho\left\{z-h(t)\right\},\\
u^{(0)}_{B_2}=\rho\left\{z-h(t)\right\}^{2};\quad v^{(0)}_{B_2}&=0;\quad w^{(0)}_{B_2}= -\rho^{2}\left\{z-h(t)\right\},\\
u^{(1)}_{B_2}= -\left\{z-h(t)\right\}^{3}+\rho^{2}\left\{z-h(t)\right\};\quad v^{(1)}_{B_2}&=a^2\left\{z-h(t)\right\};\quad w^{(1)}_{B_2}=-\rho^3+\rho\left\{z-h(t)\right\}^{2},\\
u^{(2)}_{B_2}= -\rho\left\{z-h(t)\right\}^{2};\quad v^{(2)}_{B_2}&=a^2\rho;\quad w^{(2)}_{B_2}=\rho^2\left\{z-h(t)\right\},
\end{align}
\end{subequations}
where the superscript corresponds to the $m^{\rm th}-$mode in $\vec{u}_{S}$. The $\cos m(\phi-\varphi(t))\,\left(\sin m(\phi-\varphi(t))\right)$ also gives us the corresponding $\alpha_m$ values in Eqs.~\eqref{bibc21}-~\eqref{bibc23}. The squirmer slip velocity, expressed in the form of Eq.~\eqref{eq:squimerB1}, is expanded in terms of bispherical coordinates through Eqs.~\eqref{bibc21}-~\eqref{bibc27}. For sake of clarity, we express the coefficients on the right-hand-side of Eqs.~\eqref{bibc21}-~\eqref{bibc27} in a form similar to Eq.~\eqref{eq:squimerB1}:
\begin{align}
\left\{X_n^0,Y_n^0,Z_n^0\right\}&=\frac{B_1}{a^3}\left[a\sin\vartheta(t)\left\{\left[X_n^0\right]_{B_1},\left[Y_n^0\right]_{B_1},\left[Z_n^0\right]_{B_1}\right\}\right.\nonumber\\&\left.+\beta\left(\sin^2\vartheta(t)-\frac{\cos^2\vartheta(t)}{2}\right)\left\{\left[X_n^0\right]_{B_1},\left[Y_n^0\right]_{B_1},\left[Z_n^0\right]_{B_1}\right\}\right],\\
\left\{X_n^1,Y_n^1,Z_n^1\right\}&=\frac{B_1}{a^3}\left[a\cos\vartheta(t)\left\{\left[X_n^1\right]_{B_1},\left[Y_n^1\right]_{B_1},\left[Z_n^1\right]_{B_1}\right\}+\beta\frac{\sin2\vartheta(t)}{2}\left\{\left[X_n^1\right]_{B_2},\left[Y_n^1\right]_{B_2},\left[Z_n^1\right]_{B_2}\right\}\right],\\
\left\{X_n^2,Y_n^2,Z_n^2\right\}&=\frac{B_1}{a^3}\left[\beta\frac{\cos^2\vartheta(t)}{2}\left\{\left[X_n^2\right]_{B_2},\left[Y_n^2\right]_{B_2},\left[Z_n^2\right]_{B_2}\right\}\right],
\end{align}
where 
\begin{subequations}
\begin{align}
\left[X^{0}_{n}\right]_{B_{1}}&=-\frac{2\sqrt{2}}{3}\e\sinh\eta_{0}\left[-3\cosh \eta_{0}+(2n+1)\sinh\eta_{0}\right];   \quad \alpha_0=0, \\
\left[Y^{0}_{n}\right]_{B_{1}}&=0;  \quad \alpha_0=0, \\
\left[Z^{0}_{n}\right]_{B_{1}}&=\frac{4\sqrt{2}}{3}\e\sinh\eta_{0}\left[(n^{2}+n+1)\sinh\eta_{0}-(2n+1)\cosh\eta_{0}\right];   \quad \alpha_0=0, \\
\left[X^{1}_{n}\right]_{B_{1}}&=\frac{4\sqrt{2}}{3}\e\sinh^{2}\eta_{0};    \quad \alpha_{1}=-\varphi(t),\\
\left[Y^{1}_{n}\right]_{B_{1}}&=-\frac{\sqrt{2}}{3}\e\left[3+3\cosh^2\eta_0+(2n+1)^2\sinh^2\eta_0-2(2n+1)\sinh2\eta_0\right];\nonumber\\ &\alpha_{1}=-\varphi(t),\\
\left[Z^{1}_{n}\right]_{B_{1}}&=-\frac{2\sqrt{2}}{3}\e\sinh\eta_{0}\left[(2n+1)\sinh\eta_{0}-3\cosh\eta_{0}\right];  \quad \alpha_{1}=-\varphi(t), \\
\left[X^{0}_{n}\right]_{B_{2}}&=-\frac{8\sqrt{2}}{15}\e\sinh\eta_{0}\left[(n^{2}+n+4)\cosh^{2}\eta_{0}-2(2n+1)\cosh\eta_{0}\sinh\eta_{0}-\left(n+\frac{1}{2}\right)^{2}\right];  \nonumber\\
& \alpha_0=0,\\
\left[Y^{0}_{n}\right]_{B_{2}}&=0;\quad \alpha_0=0,\\
\left[Z^{0}_{n}\right]_{B_{2}}&=\frac{2\sqrt{2}}{15}\e\sinh\eta_{0}\left[-n(1+3n+2n^2)+(6+13n+3n^2+2n^3)\cosh2\eta_0\right.\nonumber\\&\left.-3(2+3n+3n^2)\sinh2\eta_0\right];\quad\alpha_0=0,\\
\left[X^{1}_{n}\right]_{B_{2}}&=\frac{8\sqrt{2}}{15}\e\sinh^2\eta_{0}\left[(2n+1)\sinh\eta_{0}-5\cosh\eta_{0}\right];\quad \alpha_{1}=-\varphi(t),\\
\left[Y^{1}_{n}\right]_{B_{2}}&=-\frac{2\sqrt{2}}{15}\e\left[9(n^{2}+n-1)\cosh\eta_{0}-3(2+3n+3n^{2})\cosh (3\eta_{0})\right.\nonumber\\&\left.+(2n+1)\left\{15-2n-2n^{2}+2(6+n+n^{2})\cosh(2\eta_{0})\right\}\sinh\eta_{0}\right];\quad \alpha_{1}=-\varphi(t),\\
\left[Z^{1}_{n}\right]_{B_{2}}&=-\frac{2\sqrt{2}}{15}\e\sinh\eta_{0}\left[15\cosh^{2}\eta_{0}-32n\cosh \eta_{0}\sinh\eta_{0}\right.\nonumber\\&\left.+(17+8n+8n^{2})\sinh^{2}\eta_{0}-8\sinh 2\eta_{0}\right];\quad \alpha_{1}=-\varphi(t),\\
\left[X^{2}_{n}\right]_{B_{2}}&=-\frac{8\sqrt{2}}{15}\e\sinh^{3}\eta_{0};\quad \alpha_{2}=-\varphi(t),
\\
\left[Y^{2}_{n}\right]_{B_{2}}&=\frac{2\sqrt{2}}{15}\e \sinh\eta_{0}\left[15+15\cosh^{2}\eta_{0}-8\cosh\eta_{0}\sinh\eta_{0}+(2n+1)^{2}\sinh^{2}\eta_{0}\right.\nonumber\\&\left.-8n\sinh(2\eta_{0})\right];\quad \alpha_{2}=-\varphi(t),\quad{\rm and}\\
\left[Z^{2}_{n}\right]_{B_{2}}&=\frac{4\sqrt{2}}{15}\e\sinh^{2}\eta_{0}\left[-5\cosh\eta_{0}+(2n+1)\sinh\eta_{0})\right];\quad \alpha_{2}=-\varphi(t).
\end{align}
\end{subequations}

\section{Validation \label{app:validation}}
\begin{figure}[htp]
	\centering
    \includegraphics[width=\textwidth]{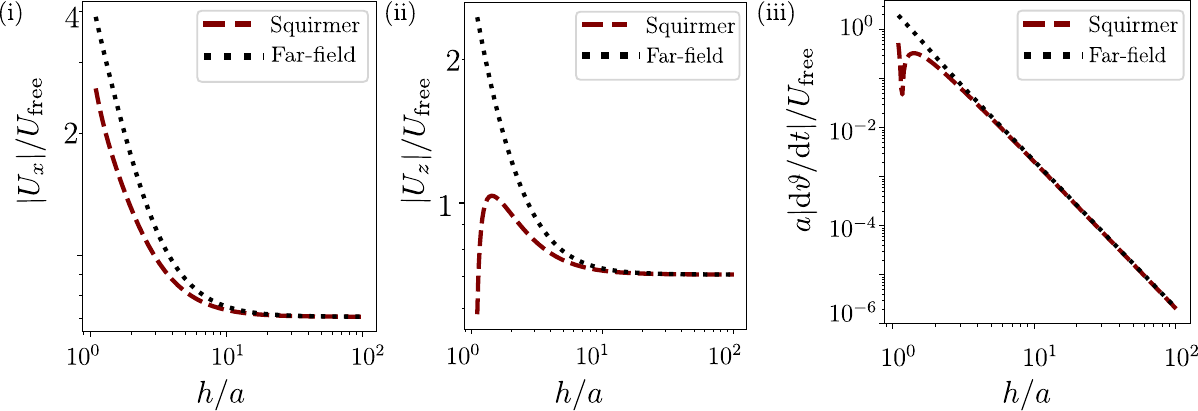}
	\caption{Comparison between velocities obtained from our bispherical approach and far-field calculations of a force-dipole near a corrugates surface~\citep{Kurzthaler:2021} as a function of swimmer-surface distance~$h/a$. We use $(x_S,y_S)=(0,0)$, $\vartheta=-0.25\pi$,  $\varphi=0$, $\beta=10$, $\lambda=2a$, and $\varepsilon=0.1$.}
	\label{fig:far-field}
\end{figure}

\subsection{Comparison to far-field results}
We validate our bispherical solution with the far-field results of a point force-dipole~\citep{Kurzthaler:2021}. Our results, see Fig.~\eqref{fig:far-field}, show that for swimmer-surface distances $h\gtrsim10a$ the bispherical solution approaches the far-field solution and agrees with it perfectly. At shorter distances, qualitative differences appear, highlighting the importance of near-field hydrodynamic flows.

\subsection{Validation with numerical simulations} 
\begin{figure}[htp]
	\centering
\includegraphics[width=\textwidth]{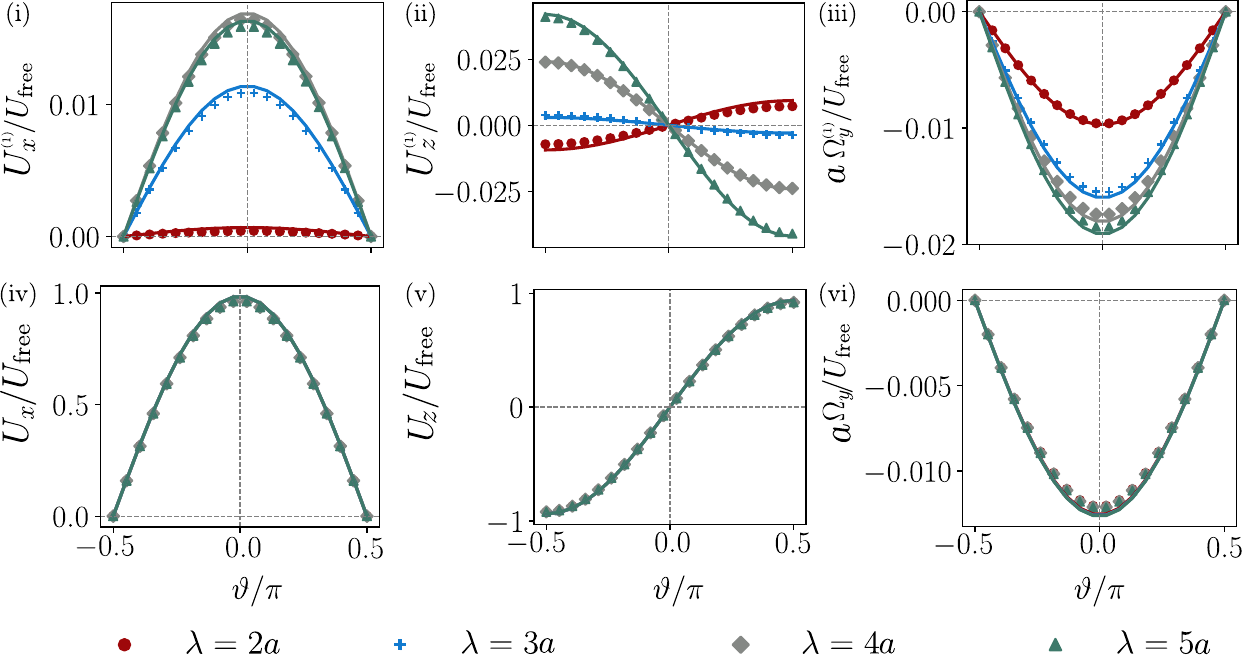}
	\caption{Comparison of (i-iii) the perturbed velocities due to surface roughness and (iv-vi) the full velocities obtained from our bispherical approach (lines) and MRS (symbols) as a function of pitch angle $\vartheta$ and for various wavelengths $\lambda$. Here, we use $\beta=0$, $\varphi=0$, and $(x_S,y_S,h)=(0,0,2a)$.}
	\label{fig:MRS}
\end{figure}
We validate our perturbative approach with numerical simulations using MRS. For small $\varepsilon=0.01$ our predictions match the numerical results quantitatively [Fig.~\eqref{fig:MRS}]. We note that for larger $\varepsilon$ differences become apparent but the results do not change qualitatively.

\section{Impact of initial distance and repulsive force \label{appC}}
\begin{figure}[htp]
	\centering
    \includegraphics[width=\textwidth]{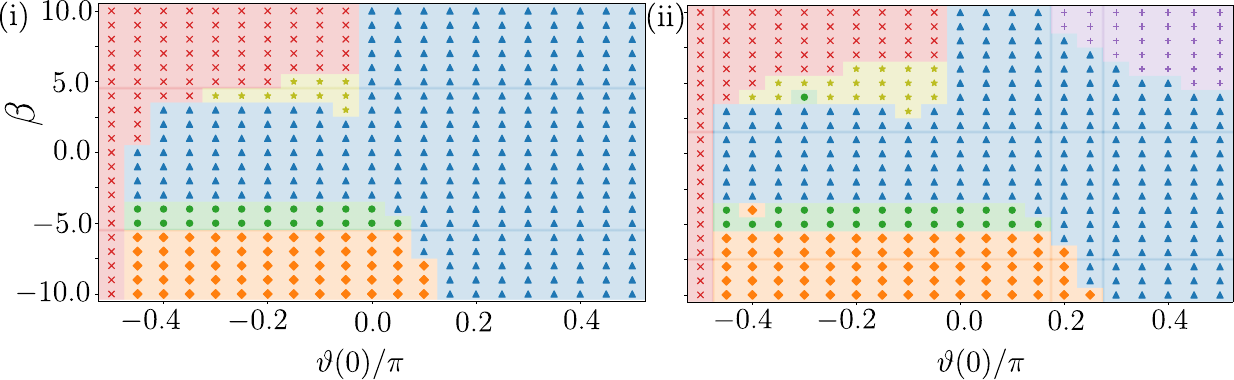}
	\caption{Phase diagram (as in Fig.~\eqref{fig:phase_plot}) with different (i)~initial distance $h(0)=3a$ and (ii) repulsive strength $A/(\mu U_{\rm free}a)=10^7$ and $B=100a$, giving a closest swimmer-surface distance of $h_{\rm min}=1.13a$. Here, we consider a wavelength of $\lambda=a$, $\varphi(0)=0$, and $\varepsilon=0.1$. We use in (i) $A/(\mu U_{\rm free}a)=10^7$ and $B=90a$ and set $h(0)=1.5 a$ in (ii).}
	\label{fig:robustness}
\end{figure}
To test the robustness of our results, we vary the initial distance $h(0)$ and the magnitude of the repulsive force (Fig.~\eqref{fig:robustness}). Both variations do not change the qualitative features of our results. We note that for $h(0)=3a$, the `trapping looking up' behavior of pullers disappears as the self-propulsion overcomes the hydrodynamic attraction by the surface. These analyses also hold for other wavelengths $\lambda=0.5a,\,2a,\,{\rm and}\,4a$.






\bibliography{biblio.bib}

\end{document}